\documentclass[aps,pra,twocolumn,showpacs,superscriptaddress,floatfix]{revtex4}
\usepackage{graphicx}
\usepackage{nicefrac}
\usepackage{amsmath}
\usepackage{amsfonts}
\usepackage{amssymb}
\usepackage{amsthm}
\usepackage{epsf}
\usepackage{bm}
\usepackage{bbm}
\usepackage{longtable}

\usepackage{dcolumn}

\newcolumntype{.}{D{x}{}{-1}}

\newcommand{\bsigma}{\vec{\sigma}}
\newcommand{\balpha}{\vec{\alpha}}
\newcommand{\bnabla}{\vec{\nabla}}
\newcommand{\bGamma}{\vec{\Gamma}}

\newcommand{\bfr}{\vec{r}}

\newcommand{\bfp}{\vec{p}}
\newcommand{\bfq}{\vec{q}}

\newcommand{\bfpp}{\vec{p}^{\prime}}

\newcommand{\hp}{\hat{\bfp}}
\newcommand{\hr}{\hat{\bfr}}

\newcommand{\Za}{{Z\alpha}}

\newcommand{\vare}{\varepsilon}

\newcommand{\pr}{^{\prime}}

\newcommand{\half}{\nicefrac12}

\newcommand{\SixJ}[6]{
        \left\{
        \begin{array}{ccc}
        #1  & #2  & #3 \\
        #4  & #5  & #6 \\
        \end{array}
        \right\}
        }

\newcommand{\lbr}{\langle} \newcommand{\rbr}{\rangle}

\begin{document}

\title{QED calculation of the nuclear magnetic shielding for hydrogen-like ions}

\author{V.~A. Yerokhin}
\affiliation{Max~Planck~Institute for Nuclear Physics, Saupfercheckweg~1, D~69117 Heidelberg, Germany}
\affiliation{ExtreMe Matter Institute
EMMI, GSI Helmholtzzentrum f\"ur Schwerionenforschung, Planckstra{\ss}e~1, D-64291 Darmstadt,
Germany}
\affiliation{St.~Petersburg State Polytechnical University, Polytekhnicheskaya~29,
        St.~Petersburg 195251, Russia}

\author{K. Pachucki} \affiliation{Faculty of
        Physics, University of Warsaw, Ho\.{z}a~69, 00--681 Warsaw, Poland}

\author{Z. Harman}
\affiliation{Max~Planck~Institute for Nuclear Physics, Saupfercheckweg~1, D~69117 Heidelberg, Germany}

\affiliation{ExtreMe Matter Institute
EMMI, GSI Helmholtzzentrum f\"ur Schwerionenforschung, Planckstra{\ss}e~1, D-64291 Darmstadt,
Germany}

\author{C.~H. Keitel}
\affiliation{Max~Planck~Institute for Nuclear Physics, Saupfercheckweg~1, D~69117 Heidelberg, Germany}

\begin{abstract}

We report an {\em ab initio} calculation of the shielding of the nuclear
magnetic moment by the bound electron in hydrogen-like ions.
This investigation takes into account several effects that have not been
calculated before (electron self-energy, vacuum
polarization, nuclear magnetization distribution), thus bringing the theory to the point where further progress is
impeded by the uncertainty due to nuclear-structure effects.
The QED corrections are
calculated to all orders in the nuclear binding strength parameter and, independently,
to the leading order in the expansion in this parameter.
The results obtained
lay the ground for the high-precision determination of nuclear magnetic dipole
moments from measurements of the $g$-factor of hydrogen-like ions.

\end{abstract}


\maketitle

\section{Introduction}

Last years showed much progress in the
experimental determination of the $g$-factors of hydrogen-like ions
\cite{haeffner:00:prl,verdu:04,sturm:11,sturm:11:b}. Measurements,
accurate up to a few parts in $10^{11}$ \cite{sturm:11:b}, were
performed by studying a single ion confined in a Penning trap.
These experiments provided the stringent tests of bound-state
quantum electrodynamics (QED) theory and yielded the best determination
of the electron mass \cite{mohr:08:rmp}.

In order to match the experimental accuracy, many sophisticated
calculations were performed during the past years, in particular
those of the one-loop self-energy
\cite{blundell:97:pra,persson:97:g,yerokhin:02:prl}, the one-loop
vacuum-polarization \cite{karshenboim:02:plb}, the nuclear recoil
\cite{shabaev:02:recprl}, and the two-loop QED effects
\cite{pachucki:04:prl,pachucki:05:gfact}. These calculations made it possible to
determine the electron mass from the experimental values of the
bound-electron $g$ factor \cite{beier:02:prl}. The
theoretical accuracy of the bound-electron $g$ factor in hydrogen-like ions
is presently at the $10^{-11}$ level \cite{pachucki:05:gfact}
for light elements up to carbon but deteriorates quickly for heavier elements
because of the unknown higher-order two-loop QED effects scaling with the
nuclear charge number $Z$ as $Z^5$.

The experimental investigations of the $g$-factors of
hydrogen-like ions were so far performed for ions with spinless nuclei.
However, when applied to the ions with a non-zero nuclear spin, such
investigations can be useful as a new method for the determination of
the magnetic dipole moments of the nuclei. This method has important
advantages over the more traditional approaches, such as nuclear
magnetic resonance (NMR), atomic beam magnetic resonance, collinear laser
spectroscopy, and optical pumping (OP). These advantages
are that (i) the simplicity of the system under investigation (a
hydrogen-like ion) allows for an {\em ab initio} theoretical description with
a reliable estimation of uncalculated effects and (ii) the influence of the
nuclear-structure effects
(which are the main limiting factors for theory) is relatively weak. This is
in contrast to the existing methods, in which the experimental data should be
corrected for several physical effects, which are difficult to calculate.
Among such effects is the diamagnetic shielding of the external magnetic field by
the electrons in the atom. The NMR results should be also corrected
for the paramagnetic chemical shift caused by the chemical environment
\cite{ramsey:50:dia} and the OP data are sensitive to the hyperfine mixing of
the energy levels \cite{lahaye:70}. Significant (and generally unknown)
uncertainties of calculations of these effects often lead to ambiguities in
the published values of nuclear magnetic moments
\cite{gustavsson:98:pra}.

The goal of the present investigation is to perform an {\it ab initio}
calculation of the $g$-factor of a hydrogen-like ion with a non-zero nuclear
spin. It can be demonstrated that the nuclear-spin dependent part of the
atomic $g$-factor can be parameterized in terms of the nuclear magnetic shielding
constant $\sigma$, which describes
the effective reduction of the coupling of the nuclear magnetic moment
$\vec{\mu}$ to an external magnetic field $\vec{B}$
caused by the shell electron(s),
\begin{eqnarray}
-\vec{\mu} \cdot \vec{B} \to -\vec{\mu} \cdot \vec{B}\, (1-\sigma) \,.
\end{eqnarray}

The relativistic theory of the $g$-factor of a hydrogen-like ion with a
non-zero nuclear spin (and, thus, the theory of the nuclear magnetic shielding)
was examined in detail in Ref.~\cite{moskovkin:04}. In the
present work, we go beyond the relativistic description of the nuclear
magnetic shielding and
calculate the dominant corrections to it, namely, the self-energy, the vacuum-polarization, and
the nuclear magnetization distribution corrections.
As a result, we bring the theory to the
point where the uncertainty due to nuclear-structure effects impedes further progress.

The main challenge of the present work is the calculation of the self-energy
correction. To the best of our knowledge, the only previous attempt to address it
was made in Ref.~\cite{rudzinski:09}. In that work, the self-energy
contribution to the shielding constant was estimated by the leading logarithm
of its $\Za$ expansion (where $\alpha$ is the
fine-structure constant). In our work, we
calculate the self-energy correction rigorously to all orders in the binding nuclear
strength parameter $\Za$  and, independently, perform an analytical
calculation to the leading order in the expansion in this parameter
(including both the logarithm and constant terms). First results of this work were reported in
Ref.~\cite{yerokhin:11:prl}.

The rest of the paper is organized as follows. In Sec.~\ref{sec:1} we
summarize the relativistic theory of the $g$-factor of an ion with a
non-zero nuclear spin and the theory of the nuclear magnetic shielding. Our calculation of
the self-energy and vacuum-polarization corrections to all orders in $\Za$ is described in
Sec.~\ref{sec:2}. In Sec.~\ref{sec:Zaexp} we report the calculation of the
QED correction to the nuclear magnetic shielding to the leading order
in the $\Za$ expansion. Sec.~\ref{sec:3} deals with the other effects,
namely, the nuclear magnetization
distribution, the nuclear recoil, and the quadrupole interaction.
Numerical results and discussion are given in
Sec.~\ref{sec:4}. The paper ends with conclusion in
Sec.~\ref{sec:5}.

The relativistic units ($m$ = $\hbar = c = 1$) and the charge units $\alpha =
e^2/(4\pi)$ are used throughout this paper.

\section{Leading-order magnetic shielding}
\label{sec:1}

We consider a hydrogen-like ion with a non-zero spin nucleus placed in a weak
homogenous magnetic field $\vec{B}$ directed along the $z$ axis. Assuming that
the energy shift due to the interaction with the $\vec{B}$ field (the Zeeman
shift) is much smaller than the hyperfine-structure splitting (hfs), the energy
shift can be expressed in terms of the $g$ factor of the atomic system $g_F$,
\begin{eqnarray} \label{eq1}
\Delta E = g_F\,\mu_0\,B\,M_F\,,
\end{eqnarray}
where $B = |\vec{B}|$, $\mu_0 = |e|/(2m)$ is the Bohr magneton, $e$ and $m$ are
the elementary charge and the electron mass, respectively,
and $M_F$ is the $z$ projection of the total angular
momentum of the system $F$. To the leading order, the energy shift is given by
\begin{eqnarray} \label{eq2}
\Delta E = \lbr FM_F|\left[-\frac{e}{2}\,(\bfr\times\balpha)\cdot\vec{B}-\vec{\mu}\cdot\vec{B}\right]|FM_F\rbr\,,
\end{eqnarray}
where $|FM_F\rbr \equiv |jIFM_F\rbr$ is the wave function of the ion, $j$ and
$I$ are the angular momentum quantum numbers of the electron and nucleus, respectively;
$\vec{\mu}$
is the operator of the magnetic moment of the nucleus.
The matrix element (\ref{eq2}) can easily be evaluated, yielding the well-known
leading-order relativistic result \cite{bethesalpeter},
\begin{eqnarray} \label{eq3}
g^{(0)}_F = g_j^{(0)}\, \frac{\lbr \vec{j}\cdot\vec{F} \rbr}{F(F+1)}
 - \frac{m}{m_p}\,g_I\, \frac{\lbr \vec{I}\cdot\vec{F} \rbr}{F(F+1)}\,,
\end{eqnarray}
where $g_j^{(0)}$ is the Dirac bound-electron $g$ factor \cite{breit:28},
$g_I = \mu/(\mu_NI)$ is the nuclear $g$ factor,
$\mu = \langle II | \vec{\mu}|II \rangle$ is the nuclear magnetic moment,
$\mu_N = |e|/(2m_p)$ is the nuclear magneton,
$m_p$ is the proton mass,
and $\lbr \vec{j}\cdot\vec{F} \rbr$ and $\lbr \vec{I}\cdot\vec{F} \rbr$ are
the angular-momentum recoupling coefficients,
\begin{eqnarray}
\lbr \vec{j}\cdot\vec{F} \rbr &=& [F(F+1)-I(I+1)+j(j+1)]/2\,,\\
\lbr \vec{I}\cdot\vec{F} \rbr &=& [F(F+1)+I(I+1)-j(j+1)]/2\,.
\end{eqnarray}

Generalizing the leading-order result to include the higher-order correction,
we write the atomic $g$ factor $g_F$ as
\begin{eqnarray} \label{eq4a}
g_F = g_j\, \frac{\lbr \vec{j}\cdot\vec{F} \rbr}{F(F+1)}
 - \frac{m}{m_p}\,g_I\, (1-\sigma)\,\frac{\lbr \vec{I}\cdot\vec{F} \rbr}{F(F+1)}\,.
\end{eqnarray}
In the above equation, the bound-electron
$g$ factor $g_j = g_j^{(0)} + \alpha/(2\pi)+ \ldots$ incorporates all
corrections that do not depend on the nuclear spin, whereas the nuclear
shielding constant $\sigma$ parameterizes the nuclear-spin dependent corrections.
The bound-electron $g$ factor $g_j$ has been extensively studied during the
last years, both theoretically
\cite{yerokhin:02:prl,karshenboim:02:plb,shabaev:02:recprl,pachucki:04:prl,pachucki:05:gfact} and experimentally
\cite{haeffner:00:prl,verdu:04,sturm:11}. The goal of the present work is the
{\em ab initio} theoretical description of the nuclear shielding parameter
$\sigma$.

It can be seen from Eq.~(\ref{eq4a}) that the nuclear-spin dependent
contribution to the atomic $g$ factor $g_F$ is suppressed by the
electron-to-proton mass ratio and thus is by about three orders of magnitude
smaller than the nuclear-spin independent part proportional to $g_j$. It is,
however, possible to form a combination of the atomic $g$ factors which is free
of the nuclear-spin independent contributions. So, for ions with the nuclear spin
$I>\nicefrac12$,
we introduce the sum of the atomic $g$ factors $\overline{g}$ that is directly
proportional to the nuclear magnetic moment,
\begin{eqnarray} \label{eq4b}
\overline{g} \equiv g_{F = I+\nicefrac12}+ g_{F = I-\nicefrac12}
   = -2\frac{m}{m_p}\frac{\mu}{\mu_NI}\,(1-\sigma)\,.
\end{eqnarray}
This combination of the $g$ factors is particularly convenient
for the determination of the nuclear magnetic dipole moments from experiment. Indeed,
if both the $g_{F = I+\nicefrac12}$ and $g_{F = I-\nicefrac12}$
$g$-factors are measured and $\sigma$ is known from theory, Eq.~(\ref{eq4b})
determines the nuclear magnetic moment $\mu$.
For the ions with a nuclear spin $I=\nicefrac12$, Eq.~(\ref{eq4b}) is not applicable
and the nuclear magnetic moment should be determined from Eq.~(\ref{eq4a}).

Contributions to the nuclear magnetic shielding are described by Feynman diagrams with
two external interactions, one with the external magnetic field (the Zeeman interaction),
\begin{eqnarray}
V_{\rm zee}(r) = \frac{|e|}{2}\,\vec{B}\cdot (\bfr\times\balpha) \,,
\end{eqnarray}
and another, with the magnetic dipole field of the nucleus (the hfs interaction)
\begin{eqnarray}
V_{\rm hfs}(r) = \frac{|e|}{4\pi}\, \vec{\mu}\cdot \frac{\bfr\times \balpha}{r^3}\,.
\end{eqnarray}
The leading-order contribution to the magnetic shielding comes from the following
energy shift
\begin{eqnarray} \label{eq5}
\Delta E = 2\,\sum_{n\ne a} \frac1{\vare_a-\vare_n}
 \lbr a|V_{\rm zee}|n\rbr\lbr n|V_{\rm hfs}|a\rbr\,,
\end{eqnarray}
where the summation runs over the whole Dirac spectrum with the reference state
excluded. As follows from Eqs.~(\ref{eq1}) and (\ref{eq4a}), contributions to
the shielding constant $\delta \sigma$  are obtained from the corresponding energy shifts
$\delta E$ by
\begin{eqnarray}
\delta \sigma = \frac{\delta E}{\mu\, B\, M_F\,
\frac{\lbr \vec{I}\cdot\vec{F} \rbr}{IF(F+1)}}\,.
\end{eqnarray}

For the electronic states with $j_a=1/2$, all nuclear
quantum numbers in Eq.~(\ref{eq5}) can be factorized out.
The expression for the shielding constant is then obtained
from that formula by using the following substitutions
\begin{eqnarray} \label{eqsubst}
& V_{\rm zee} \to \widetilde{V}_{\rm zee} \equiv (\bfr\times\balpha)_0\,, \ \ \
V_{\rm hfs} \to \widetilde{V}_{\rm hfs} \equiv \frac{\displaystyle (\bfr\times\balpha)_0}{\displaystyle r^3}\,, \ \ \
 \nonumber \\ &
| a \rbr \to | a_{\nicefrac12} \rbr \,, \ \ \
| n \rbr \to | n_{\nicefrac12} \rbr \,,
 \nonumber \\ &
2 \to \alpha \ \ \mbox{\rm (prefactor)}\,,
\end{eqnarray}
where the zero subscript refers to the zero spherical component of the vector.
Here and in what follows, $|n_{\nicefrac12}\rbr \equiv |\kappa_n,\mu_n=\nicefrac12\rbr$ denotes the Dirac state with
the relativistic angular quantum number $\kappa_n$
and the fixed momentum projection $\mu_n=1/2$.

So, for $j_a=1/2$, the leading contribution to the
magnetic shielding is given by a simple expression
\begin{align}
\sigma^{(0)} = &\,  \alpha
\sum_{n\ne a}\frac1{\vare_a-\vare_n}\,
  \lbr a_{\nicefrac12}| \widetilde{V}_{\rm zee} |n_{\nicefrac12}\rbr
  \lbr n_{\nicefrac12}| \widetilde{V}_{\rm hfs} |a_{\nicefrac12}\rbr\,.
\end{align}
Performing the angular integrations with help of
Eq.~(\ref{app:eq3}), one gets for the $ns$ reference state
\begin{eqnarray} \label{eq:leading}
\sigma^{(0)} = \alpha \sum_{\kappa_n = -1,2} x_{\kappa_n}^2\,
\sum_{n\ne a} \frac{R_{an}^{(1)}\,R_{na}^{(-2)}}{\vare_a-\vare_n}\,,
\end{eqnarray}
where $R^{(\alpha)}$ are the radial integrals of the form
\begin{eqnarray} \label{eqrad}
R^{(\alpha)}_{ab} = \int_0^{\infty}dr\,r^{2+\alpha}\bigl[g_a(r)f_b(r)+f_a(r)g_b(r) \bigr]\,,
\end{eqnarray}
$g(r)$ and $f(r)$ are the upper and lower radial components of the Dirac wave
function, respectively, and the angular prefactors $x_{\kappa_n}$ are given by
$x_{\kappa_n=-1}=-2/3$, $x_{\kappa_n=2}=-\sqrt{2}/3$.

For the point nuclear model, the sum over the Dirac spectrum in the above expression can be
evaluated analytically \cite{moore:99,pyper:99:a,pyper:99:b}, see also
more recent studies \cite{moskovkin:04,ivanov:09},
\begin{eqnarray} \label{eq5a}
\sigma^{(0)} &=& -\frac{4\alpha\Za}{9}\left(\frac13-\frac1{6(1+\gamma)}
 +\frac{2}{\gamma}-\frac3{2\gamma-1} \right)
  \nonumber \\
  &=& \alpha\Za\left(\frac13 + \frac{97}{108}(\Za)^2+\ldots \right)\,,
\end{eqnarray}
where $\gamma = \sqrt{1-(\Za)^2}$. For the extended nucleus, the
calculation is easily performed numerically \cite{moskovkin:04}.

\section{QED correction}
\label{sec:2}

\begin{figure*}
\centerline{\includegraphics[width=0.9\textwidth]{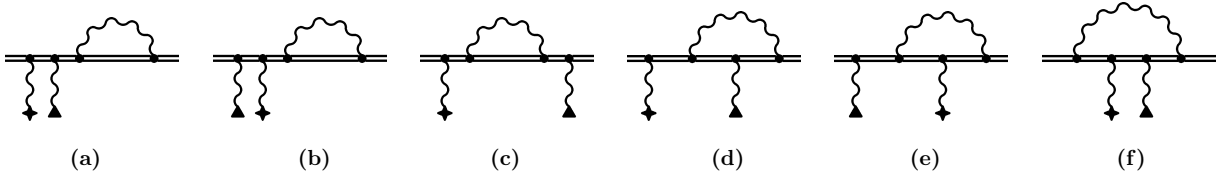}}
\caption{Self-energy correction to the nuclear magnetic shielding. Double line
  represents the electron in the binding nuclear field. The wave line terminated
  by a triangle represents the dipole hyperfine interaction with the nucleus and
  the wave line terminated by a cross represents the interaction with the
  external magnetic field.
 \label{fig:1} }
\end{figure*}

\subsection{Self-energy: General formulas}

The self-energy correction in the presence of the external magnetic field
and the magnetic dipole field of the nucleus is graphically represented by
six topologically non-equivalent
Feynman diagrams shown in Fig.~\ref{fig:1}. General expressions for these
diagrams are conveniently obtained by using the two-time Green's function method
\cite{shabaev:02:rep}. The resulting formulas are summarized below.

\subsubsection{Perturbed-orbital contribution}

The irreducible contributions of diagrams on Figs.~\ref{fig:1}(a)-(c) can be
represented in terms of the one-loop self-energy
operator $\Sigma$. This part will be termed as the perturbed-orbital
contribution. The corresponding energy shift is
\begin{eqnarray} \label{po1}
\Delta E_{\rm po} &=&
 2\, \lbr \Sigma_R \,\frac1{(\vare_a-H)^{\prime}}\, V_{\rm zee}\,
                 \frac1{(\vare_a-H)^{\prime}}\, V_{\rm hfs} \rbr
\nonumber \\ &&
+ 2\, \lbr \Sigma_R \,\frac1{(\vare_a-H)^{\prime}}\, V_{\rm hfs}\,
                 \frac1{(\vare_a-H)^{\prime}}\, V_{\rm zee} \rbr
\nonumber \\ &&
-2\, \lbr \Sigma_R \,\frac1{(\vare_a-H)^{2^{\prime}}}\, V_{\rm zee}\rbr
                 \lbr V_{\rm hfs} \rbr
\nonumber \\ &&
-2\, \lbr \Sigma_R \,\frac1{(\vare_a-H)^{2^{\prime}}}\, V_{\rm hfs}\rbr
                 \lbr V_{\rm zee} \rbr
\nonumber \\ &&
-2\, \lbr \Sigma_R \rbr
\lbr V_{\rm zee} \,\frac1{(\vare_a-H)^{2^{\prime}}}\, V_{\rm hfs}\rbr
\nonumber \\ &&
+2\, \lbr V_{\rm zee} \,\frac1{(\vare_a-H)^{\prime}}\, \Sigma_R\,
                 \frac1{(\vare_a-H)^{\prime}}\, V_{\rm hfs} \rbr\,,
\nonumber \\ &&
\end{eqnarray}
where we used the short-hand notations for the reduced Green's function,
\begin{equation}
\frac1{(\vare_a-H)^{{\prime}}} = \sum_{n\ne a}
 \frac{|n\rbr\lbr n|}{\vare_a-\vare_n}\,,
\end{equation}
and its derivative
\begin{equation}
\frac1{(\vare_a-H)^{2^{\prime}}} = \sum_{n\ne a}
 \frac{|n\rbr\lbr n|}{(\vare_a-\vare_n)^2}\,.
\end{equation}
The (unrenormalized) self-energy operator $\Sigma(\vare)$ is defined by its matrix elements as follows
\begin{eqnarray}
\lbr i | \Sigma(\vare)|k\rbr &=& \frac{i}{2\pi}\int_{-\infty}^{\infty} d\omega
  \sum_{n}
  \frac{\lbr in|I(\omega)|nk\rbr}{\vare-\omega-u\,\vare_n} \,,
\end{eqnarray}
where $I(\omega)$ is the operator of the electron-electron interaction,
$I(\omega) = e^2\alpha_{\mu}\alpha_{\nu}D^{\mu\nu}(\omega)$,
$D^{\mu\nu}(\omega)$ is the photon propagator, $\alpha_{\mu}$ are the
Dirac matrices, and $u \equiv
1-i0$, where $i0$ is a small imaginary addition that
defines the positions of the poles of
the electron propagators with respect to the integration contour.
The summation over $n$ runs over the complete Dirac spectrum.
Renormalization of the one-loop self-energy operator is well described in the
literature, see, e.g., Ref.~\cite{snyderman:91}. In this work, the
renormalized part of the self-energy operator is defined as
\begin{align}
\Sigma_R(\vare) = \Sigma(\vare)-\beta\,\delta m - (\vare-\balpha\cdot\bfp - \,V_C
-\beta m)\,B^{(1)}\,,
\end{align}
where $\delta m$ is the mass counterterm, $\beta$ is the Dirac $\beta$ matrix, $B^{(1)}$ is the
one-loop renormalization constant, $V_C$ is the binding Coulomb potential
of the nucleus, and the renormalization is to be performed
in momentum space with a covariant regularization of ultraviolet (UV) divergencies.
Details on the renormalization procedure and explicit formulas
for $\Sigma_R(\vare)$ can be found in  Refs.~\cite{yerokhin:99:pra,yerokhin:03:epjd}.

\subsubsection{Single-vertex contributions}

The irreducible contribution of the diagram shown in Fig.~\ref{fig:1}(d),
together with the corresponding derivative term,
is referred to as the {\em hfs-vertex} contribution.
It is given by
\begin{eqnarray} \label{vrhfs}
\Delta E_{\rm vr,hfs} &=& 2\, \lbr \Gamma_{{\rm hfs},R}  \,\frac1{(\vare_a-H)^{{\prime}}}\,
V_{\rm zee}\rbr
 \nonumber \\ &&
+ 2\,\lbr \Sigma^{\prime}_R  \,\frac1{(\vare_a-H)^{{\prime}}}\,
V_{\rm zee}\rbr\,\lbr V_{\rm hfs}\rbr\, ,
\end{eqnarray}
where $\Gamma_{{\rm hfs},R}\equiv \Gamma_{{\rm hfs},R}(\vare_a)$ is the
renormalized part of the 3-point vertex
representing the interaction with the hfs field
and $\Sigma^{\prime}_R\equiv
\Sigma^{\prime}_R(\vare_a)$ is the derivative of the
renormalized self-energy operator over the energy argument,
$\Sigma^{\prime}_R(\vare_a) =
\left. d/(d\vare)\Sigma_R(\vare)\right|_{\vare= \vare_a}$.
The unrenormalized 3-point hfs vertex operator is defined by its matrix elements as
\begin{align}
\lbr i | \Gamma_{\rm hfs}(\vare)|k\rbr &\ = \frac{i}{2\pi}\int_{-\infty}^{\infty} d\omega
  \nonumber \\ & \times
  \sum_{n_1n_2}
  \frac{\lbr in_2|I(\omega)|n_1k\rbr \lbr n_1|V_{\rm hfs}|n_2\rbr }
 {(\vare-\omega-u\vare_{n_1})(\vare-\omega-u\vare_{n_2})} \,.
\end{align}
The renormalized part of the operator is obtained as
\begin{eqnarray}
\Gamma_{{\rm hfs},R}(\vare) = \Gamma_{{\rm hfs}}(\vare) -
 V_{\rm hfs} \, L^{(1)} \,,
\end{eqnarray}
where $L^{(1)}$ is the one-loop renormalization constant and the
renormalization is to be performed in momentum space with a covariant
regularization of UV divergencies, see
Ref.~\cite{yerokhin:03:epjd} for details.

The irreducible contribution of the diagram shown in Fig.~\ref{fig:1}(e),
together with the corresponding derivative term,
is referred to as the {\em Zeeman-vertex} contribution. It
is given by
\begin{eqnarray}  \label{vrzee}
\Delta E_{\rm vr,zee} &=& 2\, \lbr \Gamma_{{\rm zee},R}  \,\frac1{(\vare_a-H)^{{\prime}}}\,
V_{\rm hfs}\rbr
 \nonumber \\ &&
+ 2\,\lbr \Sigma^{\prime}_R  \,\frac1{(\vare_a-H)^{{\prime}}}\,
V_{\rm hfs}\rbr\,\lbr V_{\rm zee}\rbr\,,
\end{eqnarray}
where the 3-point Zeeman vertex is defined analogously to the
hfs one.

\subsubsection{Double-vertex contribution}

The contribution of the diagram shown on Fig.~\ref{fig:1}(f),
together with the corresponding derivative terms, will be termed as the {\em
  double-vertex} contribution. It is defined by
\begin{align} \label{dvr}
&\Delta E_{\rm d.vr}  =
 2\,\lbr \Lambda(\vare_a)\rbr
\nonumber \\ &
  + \lbr \Sigma^{\prime\prime}\rbr  \lbr V_{\rm zee}\rbr \lbr V_{\rm hfs}\rbr
 + \lbr \Gamma_{\rm hfs}^{\prime}\rbr \lbr V_{\rm zee}\rbr
 + \lbr \Gamma_{\rm zee}^{\prime}\rbr\lbr V_{\rm hfs} \rbr
\nonumber \\ &
 - 2\, \lbr V_{\rm zee} \frac1{(\vare_a-H)^{\prime}} V_{\rm hfs}\rbr\,
\frac{i}{2\pi}\int_{-\infty}^{\infty} d\omega
  \sum_{a'}
  \frac{\lbr aa'|I(\omega)|a'a\rbr}{(-\omega+i0)^2} \,,
\end{align}
where $\Lambda$ denotes the 4-point vertex
representing the interaction with both the Zeeman and hfs interactions,
\begin{align}
&\lbr i | \Lambda(\vare)|k\rbr = \frac{i}{2\pi}\int_{-\infty}^{\infty} d\omega
\nonumber \\ &\times
  \sum_{n_1n_2n_3}
  \frac{\lbr in_3|I(\omega)|n_1k\rbr \lbr n_1|V_{\rm zee}|n_2\rbr \lbr
    n_2|V_{\rm hfs}|n_3\rbr }
 {(\vare-\omega-u\vare_{n_1})(\vare-\omega-u\vare_{n_2})(\vare-\omega-u\vare_{n_3})} \,,
\end{align}
$\Sigma^{\prime\prime}\equiv\Sigma^{\prime\prime}(\vare_a)$ denotes the second derivative of the self-energy
operator over the energy argument,
$\Sigma^{\prime\prime}(\vare_a) =
\left. d^2/(d^2\vare)\,\Sigma(\vare)\right|_{\vare= \vare_a}$,
$\Gamma^{\prime}\equiv \Gamma^{\prime}(\vare_a)$ denotes the derivative of the vertex operator over the
energy argument, and $a'$ denotes the intermediate electron states with the energy
$\vare_{a'} = \vare_{a}$.
The last term in Eq.~(\ref{dvr}) is
added artifically, in order to make the whole expression for $\Delta E_{\rm
  d.vr}$ infrared (IR)
finite. The same term will be subtracted from the derivative contribution
defined below, see Eq.~(\ref{der}).

We note that all terms in Eq.~(\ref{dvr}) are UV finite, so that
there is no need for any UV regularization. There
are, however, IR divergences, which appear when
the energy of the intermediate electron states in the electron propagators
coincides with the energy of the reference state. The divergences
cancel out in the sum of individual terms in Eq.~(\ref{dvr}).

\subsubsection{Derivative contribution}

Finally, the remaining contribution will be termed as the {\em derivative}
term. It is given by
\begin{eqnarray} \label{der}
\Delta E_{\rm der} &=&
2\, \biggl[\lbr \Sigma^{\prime}_R \rbr
+\frac{i}{2\pi}\int_{-\infty}^{\infty} d\omega
  \sum_{a'}
  \frac{\lbr aa'|I(\omega)|a'a\rbr}{(-\omega+i0)^2}
\biggr]
 \nonumber \\ && \times
\lbr V_{\rm zee} \,\frac1{(\vare_a-H)^{{\prime}}}\, V_{\rm hfs}\rbr \,.
\end{eqnarray}
The second term in the brackets is added artifically, in order to compensate
the IR reference-state divergence present in the first term, making the total
expression for $\Delta E_{\rm der}$
IR finite. Note that this term is exactly the same as the
one added to Eq.~(\ref{dvr}) but has the opposite sign.

Finally, the total self-energy correction is given by the sum of the
contributions discussed above,
\begin{align} \label{eq7}
\Delta E_{\rm SE} = \Delta E_{\rm po}+ \Delta E_{\rm
  vr, hfs}+ \Delta E_{\rm  vr, zee} + \Delta E_{\rm  d.vr} + \Delta E_{\rm
  der}\,,
\end{align}
which are given by Eqs.~(\ref{po1}), (\ref{vrhfs}),
(\ref{vrzee}), (\ref{dvr}), and (\ref{der}), respectively.

\subsection{Self-energy: Calculation}

The general formulas reported so far represent contributions to the energy
shift. We now have to separate out the nuclear degrees of freedom and convert
the corrections to the energy into corrections to the shielding constant. In most cases,
this is achieved simply by using the substitutions (\ref{eqsubst}). The
double-vertex contribution, however, requires an explicit angular-momentum
algebra calculation for the separation of the nuclear variables.

We will see that most of the corrections to the shielding constant can be
regarded as generalizations of the corrections already discussed in the
literature. So, our present calculation will be largely based on the previous
investigations of the self-energy correction to the Lamb shift
\cite{yerokhin:99:pra,yerokhin:05:se}, to the
hyperfine structure \cite{yerokhin:05:hfs,yerokhin:08:prl,yerokhin:10:sehfs}, and to the $g$ factor
\cite{yerokhin:04,yerokhin:08:prl,yerokhin:10:sehfs}. The double-vertex correction of the kind similar to that
in the present work appeared in the evaluation
of the self-energy correction to the parity-nonconserving transitions in
Refs.~\cite{shabaev:05:prl,shabaev:05:pra}. However, in this work we develop a different scheme for
the evaluation of the double-vertex contribution based on the analytical representation of
the Dirac-Coulomb Green function.

\subsubsection{Perturbed-orbital contributions}

The matrix elements of the self-energy operator are diagonal in the relativistic
angular momentum quantum number $\kappa$ and the momentum projection
$\mu$. Because of this, the angular reduction of the perturbed-orbital
contribution is achieved by the same set of
substitutions  (\ref{eqsubst}) as for the leading-order magnetic
shielding. The resulting contribution to
the shielding constant is conveniently represented as
\begin{eqnarray}
\Delta \sigma_{\rm po} &=& \alpha\, \lbr a_{\nicefrac12}|\Sigma_R(\vare_a) |\delta^{(2)}a_{\nicefrac12}\rbr
 \nonumber \\ &&
 + \alpha\, \lbr \delta^{(1)}_{\rm hfs}a_{\nicefrac12}|\Sigma_R(\vare_a) |\delta^{(1)}_{\rm zee}a_{\nicefrac12}\rbr\,,
\end{eqnarray}
where the first-order perturbations of the reference-state wave function are
given by
\begin{align}
|\delta^{(1)}_{\rm hfs} a_{\nicefrac12}\rbr = \sum_{n\ne a} |n_{\nicefrac12}\rbr
 \frac1{\vare_a-\vare_{n}}\,
\lbr n_{\nicefrac12}|\widetilde{V}_{\rm hfs} |a_{\nicefrac12}\rbr \,,
\end{align}
and
\begin{align}
|\delta^{(1)}_{\rm zee} a_{\nicefrac12}\rbr = \sum_{n\ne a} |n_{\nicefrac12}\rbr
 \frac1{\vare_a-\vare_{n}}\,
\lbr n_{\nicefrac12}|\widetilde{V}_{\rm zee}|a_{\nicefrac12}\rbr \,,
\end{align}
and $|\delta^{(2)}a\rangle$ is the standard second-order perturbation
\cite{landau:III} of the
reference-state wave function induced by {\em both} interactions,
$\widetilde{V}_{\rm hfs}$ and $\widetilde{V}_{\rm zee}$.
Note that only the diagonal in $\kappa$ part of the perturbed wave
functions contributes to $\Delta \sigma_{\rm po}$.
The calculation of the non-diagonal matrix elements of the self-energy operator
is performed by a straightforward generalization of the method developed in
Ref.~\cite{yerokhin:05:se} for the first-order self-energy correction to the
Lamb shift.

\subsubsection{Hfs-vertex contributions}

The hfs-vertex correction to the energy shift (\ref{vrhfs}) for the reference
state with
$j_a=\nicefrac12$ can be converted to the correction to the magnetic shielding
by the substitution (\ref{eqsubst}). The result is
\begin{align} \label{eqhfs}
& \Delta \sigma_{\rm vr,hfs} =
\frac{i\alpha}{2\pi}\int_{-\infty}^{\infty} d\omega
 \nonumber \\ & \times
  \sum_{n_1n_2}
  \frac{\lbr a_{\nicefrac12}n_2|I(\omega)|n_1\,\delta^{(1)}_{\rm zee} a_{\nicefrac12}\rbr
 \lbr n_1|\widetilde{V}_{\rm hfs}|n_2\rbr }
 {(\vare_a-\omega-u\,\vare_{n_1})(\vare_a-\omega-u\,\vare_{n_2})}
 \nonumber \\ &
- \lbr\widetilde{V}_{\rm hfs}\rbr \,
 \frac{i\alpha}{2\pi}\int_{-\infty}^{\infty} d\omega
  \sum_{n}
  \frac{\lbr a_{\nicefrac12}n|I(\omega)|n\,\delta^{(1)}_{\rm zee} a_{\nicefrac12}\rbr }
 {(\vare_a-\omega-u\,\vare_{n})^2} \,,
\end{align}
where the covariant regularization of the ultraviolet (UV) divergences is
implicitly assumed. The right-hand-side of Eq.~(\ref{eqhfs}) differs from
the vertex and reducible parts of the self-energy
correction to the hyperfine structure only by the
perturbed wave function $|\delta^{(1)}_{\rm zee} a\rbr$
in place of one of the reference-state wave functions $|a\rbr$.
The main complication brought by this difference is that the perturbed wave function contains
components with different values of the relativistic angular quantum number
$\kappa$. So, for the reference state with $\kappa_a = -1$, the perturbed wave
function has components with $\kappa = -1$ and $\kappa = 2$, both of which
contribute to the first term in Eq.~(\ref{eqhfs}), denoted in the following
as $\Delta \sigma_{\rm ver,hfs}$. The second (reducible) term contains only
the $\kappa = \kappa_a$ component of the perturbed wave function and its
calculation is done exactly as described in Ref.~\cite{yerokhin:05:hfs}.
Below, we present the generalization of formulas derived in
Ref.~\cite{yerokhin:05:hfs,yerokhin:10:sehfs} needed for the evaluation of
$\Delta \sigma_{\rm ver,hfs}$.

As explained in Ref.~\cite{yerokhin:05:hfs}, the covariant separation of the
UV divergences is conveniently performed by dividing the vertex
contribution into the zero- and many-potential parts, according to the number of
interactions with the binding Coulomb field in the electron propagator,
\begin{align}
\Delta \sigma_{\rm ver,hfs} = \Delta \sigma_{\rm ver,hfs}^{(0)} + \Delta \sigma_{\rm ver,hfs}^{(1+)}\,.
\end{align}
The zero-potential part is calculated in momentum space with the dimensional
regularization of the UV divergences. The Fourier transform of the
$\widetilde{V}_{\rm hfs}$ is done by
\begin{eqnarray}
 \frac{(\bfr \times \balpha)_0}{r^3} \to (-4\pi i)\, \frac{(\bfq \times \balpha)_0}{\bfq^2}\,,
\end{eqnarray}
where $\bfq = \bfp_1-\bfp_2$ is the transferred momentum.
The contribution of the zero-potential hfs vertex part to the shielding constant is
\begin{align} \label{e6}
\Delta \sigma^{(0)}_{\rm ver,hfs} &\ = -4\pi i \alpha
   \int \frac{d\bfp_1}{(2\pi)^3}\, \frac{d\bfp_2}{(2\pi)^3}\,
\nonumber \\ & \times
     \overline{\psi}_{a_{\nicefrac12}}(\bfp_1)\,
          \frac{\left[ \bfq \times {\vec \Gamma}_R(p_1,p_2)\right]_0}{\bfq^2}\,
       \psi_{\delta a_{\nicefrac12}}(\bfp_2)\,,
\end{align}
where
$p_1$ and $p_2$ are 4-vectors with the fixed time
component $p_1 = (\vare_a,\bfp_1)$, $p_2 = (\vare_a,\bfp_2)$, $\psi_a$ and
$\psi_{\delta a}$ are the
reference-state and the perturbed wave functions, respectively,
$\overline{\psi} = \psi^{\dag}\gamma^0$ is the Dirac conjugation,
and ${\vec
 \Gamma}_R$ is the renormalized one-loop vertex operator \cite{yerokhin:99:pra}.
For evaluating the integrals over the angular variables, it is
convenient to use the following representation of the
vertex operator sandwiched between the Dirac wave functions
\begin{align} \label{e7}
\overline{\psi}_a(\bfp_1)&\ {\vec\Gamma}_R(p_1,p_2)\,
        \psi_b(\bfp_2)
         = \frac{\alpha}{4\pi}
\left[
    {\cal R}_1 \chi^{\dag}_{\kappa_a \mu_a}(\hat{\bfp}_1)
        \, {\bsigma}\, \chi_{-\kappa_b\mu_b}(\hat{\bfp}_2)
        \right. \nonumber \\ &
   +{\cal R}_2 \chi^{\dag}_{-\kappa_a \mu_a}(\hat{\bfp}_1)
        \,{\bsigma}\, \chi_{\kappa_b\mu_b}(\hat{\bfp}_2)
        \nonumber \\ &
+ ({\cal R}_3 \,\bfp_1+ {\cal R}_4\, \bfp_2)
         \chi^{\dag}_{\kappa_a \mu_a}(\hat{\bfp}_1)
         \chi_{\kappa_b\mu_b}(\hat{\bfp}_2)
 \nonumber \\&
 + \left. ({\cal R}_5 \,\bfp_1 +{\cal R}_6\, \bfp_2)
         \chi^{\dag}_{-\kappa_a \mu_a}(\hat{\bfp}_1)
         \chi_{-\kappa_b\mu_b}(\hat{\bfp}_2) \right] \,,
\end{align}
where $\hp \equiv \bfp/|\bfp|$, $\chi_{\kappa\mu}(\hp)$ are the
spin-angular Dirac spinors \cite{rose:61},
and the scalar functions ${\cal R}_i$ are
given by Eqs.~(A7)--(A12) of Ref.~\cite{yerokhin:99:sescr}.
Integration over the angular
variables yields (cf.~Eq.~(30) of Ref.~\cite{yerokhin:10:sehfs})
\begin{widetext}
\begin{align} \label{e8}
\Delta \sigma^{(0)}_{\rm ver,hfs} &\ =
   -\frac{\alpha^2}{48\pi^5}\,  \sum_{\kappa_{\delta a}} x_{\kappa_{\delta
       a}}\, i^{l_a-l_{\delta a}}\,
      \int_0^{\infty}dp_{1r}\,dp_{2r}\int_{|p_{1r}-p_{2r}|}^{p_{1r}+p_{2r}}dq_r\,
           \frac{p_{1r}p_{2r}}{q_r}\,
\nonumber \\ & \times
     \Bigl\{
  {\cal R}_1 \, [p_{1r}K_1(\kappa_a,-\kappa_{\delta a})-p_{2r}K_1^{\prime}(\kappa_a,-\kappa_{\delta a})]
 +{\cal R}_2 \, [p_{1r}K_1(-\kappa_a,\kappa_{\delta a})-p_{2r}K_1^{\prime}(-\kappa_a,\kappa_{\delta a})]
\nonumber \\ &
        + p_{1r}p_{2r}({\cal R}_3+{\cal R}_4)\, K_2(\kappa_a,\kappa_{\delta a})\,
        + p_{1r}p_{2r}({\cal R}_5+{\cal R}_6)\, K_2(-\kappa_a,-\kappa_{\delta
          a})
\Bigr\}\,,
\end{align}
\end{widetext}
where $p_{i_r} = |\bfp_i|$, $q_r = |\bfq|$, $\kappa_{\delta a}$ is the relativistic angular
quantum number of the perturbed wave function, $l_n = |\kappa_{n}+\nicefrac12|-\nicefrac12$,
$x_{\kappa=-1}=-2/3$, $x_{\kappa=2}=-\sqrt{2}/3$,
and the basic angular integrals
$K_i(\kappa,\kappa')$ are defined and evaluated in
Appendix~\ref{app:angular}.

The many-potential vertex contribution is free from UV
divergences and thus can be calculated in coordinate space.
The result after the integration over the angular variables is
\begin{align}    \label{eq00}
\Delta \sigma_{\rm ver, hfs}^{(1+)} &\ =
 \sum_{\kappa_{\delta a} = -1,2} x_{\kappa_{\delta a}}\,
\frac{i\alpha^2}{2\pi} \int_{-\infty}^{\infty} d\omega
  \nonumber \\& \times
  \sum_{n_1n_2}
  \frac{R^{(-2)}_{n_1n_2} \, \sum_J  X_J(\kappa_1,\kappa_2)\, R_J(\omega,an_2n_1\delta a)}
 {(\vare-\omega-u\,\vare_{n_1})(\vare-\omega-u\,\vare_{n_2})}
  \nonumber \\&
-  \ \mbox{\rm subtraction}\,,
\end{align}
where $R^{(-2)}_{n_1n_2}$ is the radial integral of the hfs type given by
Eq.~(\ref{eqrad}), $X_J$ is the angular coefficient,
\begin{align}
X_J(\kappa_1,\kappa_2) =&\
\frac{(-1)^{j_{\delta a}-1/2}}{\sqrt{2}}
 \SixJ{j_1}{j_2}{1}{j_{\delta a}}{j_a}{J}
  \nonumber \\ & \times
\frac{-\kappa_1-\kappa_2}{\sqrt{3}} C_1(-\kappa_2,\kappa_1)\,\,,
\end{align}
$C_1(\kappa_a,\kappa_b)$ is the reduced matrix element of the
normalized spherical harmonics given by Eq.~(C10) of
Ref.~\cite{yerokhin:99:pra},
$R_J$ is the relativistic generalization of the Slater radial integral given by
Eqs.~(C1)--(C9) of Ref.~\cite{yerokhin:99:pra}, and the subtraction in the
last line of Eq.~(\ref{eq00}) means that the
contribution of the free propagators (already accounted for by the
zero-potential term) needs to be subtracted.

\subsubsection{Zeeman-vertex contribution}

The Zeeman-vertex correction to the energy shift (\ref{vrzee}) for
the reference state with
$j_a=\nicefrac12$ can be converted to the correction to the magnetic shielding
by the substitution (\ref{eqsubst}). The result has the form analogous to that
for the hfs-vertex contribution,
\begin{align}
& \Delta \sigma_{\rm vr,zee} =
\frac{i\alpha}{2\pi}\int_{-\infty}^{\infty} d\omega
 \nonumber \\ & \times
  \sum_{n_1n_2}
  \frac{\lbr a_{\nicefrac12}n_2|I(\omega)|n_1\,\delta^{(1)}_{\rm hfs} a_{\nicefrac12}\rbr \lbr n_1|\widetilde{V}_{\rm zee}|n_2\rbr }
 {(\vare_a-\omega-u\,\vare_{n_1})(\vare_a-\omega-u\,\vare_{n_2})}
 \nonumber \\ &
- \lbr \widetilde{V}_{\rm zee}\rbr \,
 \frac{i\alpha}{2\pi}\int_{-\infty}^{\infty} d\omega
  \sum_{n}
  \frac{\lbr a_{\nicefrac12}n|I(\omega)|n\,\delta^{(1)}_{\rm hfs} a_{\nicefrac12}\rbr }
 {(\vare_a-\omega-u\,\vare_{n})^2} \,,
\end{align}
where the covariant regularization of the UV divergences is
implicitly assumed. The above expression looks very similar to
Eq.~(\ref{eqhfs}) and can be evaluated almost in the same way, except for the
zero-potential vertex contribution. The expression for the zero-potential
Zeeman vertex contribution is different from the hfs case because the
momentum representation of the interaction with a constant magnetic field involves a
$\delta$-function. The diagonal matrix element of the Zeeman vertex operator was
evaluated previously in Ref.~\cite{yerokhin:04}; here we present the
generalization of the formulas required for the non-diagonal case.

The Fourier transform of the interaction with
the external magnetic field is given by
\begin{eqnarray}
\vec{B}\times\bfr \to -i(2\pi)^3\, \vec{B} \times \bnabla_{\bfp^{\prime}}
         \delta^3(\bfp-\bfp^{\prime})\,.
\end{eqnarray}
The contribution of the zero-potential vertex to the shielding constant is
\begin{eqnarray}
\Delta \sigma_{\rm ver,zee}^{(0)} &=& -i\alpha
    \int \frac{d\bfp\, d\bfpp}{(2\pi)^3}\, \overline{\psi}_{a_{\nicefrac12}}(\bfp)
\nonumber \\ && \times
    \left[\bnabla_{\bfpp} \delta^3(\bfp-\bfpp)
      \times \bGamma_R(p,p\pr) \right]_0 \psi_{\delta a_{\nicefrac12}}(\bfpp)\,,
\nonumber \\
\end{eqnarray}
where the gradient $\bnabla_{\bfpp}$ acts on the $\delta$-function only.
This expression is transformed by integrating by parts and
carrying out the integration with the $\delta$-function analytically.
The result after the angular integration
consists of two parts (cf. Eqs.~(27) and (36) of Ref.~\cite{yerokhin:04}),
\begin{equation}
\Delta \sigma_{\rm ver,zee}^{(0)}  = \Delta \sigma_{\rm ver,zee,1}^{(0)}
+ \Delta \sigma_{\rm ver,zee,2}^{(0)} \,,
\end{equation}
where
\begin{align}
\Delta \sigma_{\rm ver,zee,1}^{(0)}  &\ =
 \frac{\alpha^2}{\pi} \sum_{\kappa_{\delta a}} x_{\kappa_{\delta
     a}}\,i^{l_a-l_{\delta a}}\,
\int_0^{\infty} \frac{p^2dp}{8\pi^3}\, A(\rho)\,
\nonumber \\ & \times
  \biggl[ g_a \tilde{g}_{\delta a}\,A_1(\kappa_a,\kappa_{\delta a})
         +f_a \tilde{f}_{\delta a}\,A_1(-\kappa_a,-\kappa_{\delta a})
\nonumber \\ &
 -p\, g_af_{\delta a}\,A_2(\kappa_a,-\kappa_{\delta a})
 -p\, f_ag_{\delta a}\,A_2(-\kappa_a,\kappa_{\delta a})\biggr]\,,
\end{align}
and
\begin{align}
  &\ \Delta \sigma_{\rm ver,zee,2}^{(0)} =
 -\frac{\alpha^2}{4\pi} \sum_{\kappa_{\delta a}} x_{\kappa_{\delta a}}\, i^{l_a-l_{\delta a}}\,
\int_0^{\infty} \frac{p^2dp}{8\pi^3}\,
\nonumber \\ & \times
    \biggl\{ b_1(\rho)\, \bigl[ g_af^{\prime}_{\delta a} A_2(\kappa_a,-\kappa_{\delta a})
  + f_ag^{\prime}_{\delta a} A_2(-\kappa_a,\kappa_{\delta a})\bigr]
\nonumber \\ &
 + \frac1p\, b_1(\rho)\, \bigl[ g_af_{\delta a} A_3(\kappa_a,-\kappa_{\delta a})
  + f_ag_{\delta a} A_3(-\kappa_a,\kappa_{\delta a})\bigr]
\nonumber \\ &
 + b_2(\rho)\, \bigl[ \tilde{g}_ag_{\delta a} A_4(\kappa_a,\kappa_{\delta a})
  + \tilde{f}_af_{\delta a} A_4(-\kappa_a,-\kappa_{\delta a})\bigr]
\nonumber \\ &
 + b_3(\rho)\, \bigl[ g_ag_{\delta a} A_4(\kappa_a,\kappa_{\delta a})
  - f_af_{\delta a} A_4(-\kappa_a,-\kappa_{\delta a})\bigr]\biggr\}\,,
\end{align}
where $\tilde{g}_{\delta a} = \vare_a g_{\delta a}+p f_{\delta a}$,
$\tilde{f}_{\delta a} = \vare_a f_{\delta a}+p g_{\delta a}$,
$\tilde{g}_{a} = \vare_a g_{a}+p f_{a}$,
$\tilde{f}_{a} = \vare_a f_{a}+p g_{a}$,
$g^{\prime} = dg(p)/dp$ and $f^{\prime} = df(p)/dp$,
the scalar functions $A(\rho)$, $b_i(\rho)$ are given by Eqs.~(24), (30)-(32)
of Ref.~\cite{yerokhin:04} and $A_i(\kappa_1,\kappa_2)$ are the basic angular integrals
defined and evaluated in Appendix~\ref{app:angular}.

Note that the expression for $\Delta \sigma_{\rm ver,zee,2}^{(0)}$ is
non-symmetric with respect to $a$ and $\delta a$, but
the result does not change when the two wave functions are interchanged.

\subsubsection{Double-vertex contribution}

The double-vertex correction is defined by Eq.~(\ref{dvr}).
All parts of it are UV finite and thus can be evaluated in coordinate
space. We denote the individual
terms in the right-hand-side of Eq.~(\ref{dvr}) by $\delta_iE$ and
consider each of them separately,
\begin{eqnarray}
\Delta E_{\rm d.vr} = \sum_{i=1}^5 \delta_iE\,.
\end{eqnarray}
The first term is
\begin{align}
& \delta_1E \equiv 2\,\lbr \Lambda \rbr = 2\, \frac{i}{2\pi}\int_{-\infty}^{\infty} d\omega
  \nonumber \\ & \times
  \sum_{n_1n_2n_3}
  \frac{\lbr a n_3|I(\omega)|n_1 a\rbr \lbr n_1|V_{\rm zee}|n_2\rbr \lbr n_2|V_{\rm hfs}|n_3\rbr }
 {(\vare-\omega-u\,\vare_{n_1})(\vare-\omega-u\,\vare_{n_2})(\vare-\omega-u\,\vare_{n_3})} \,.
\end{align}
After separating the nuclear variables and integrating over the angles,
the contribution to the magnetic shielding is (for $j_a = 1/2$)
\begin{align}\label{delta1}
\delta_1 \sigma  &\ =
\frac{i\alpha^2}{2\pi} \int_{-\infty}^{\infty} d\omega
  \sum_{n_1n_2n_3}
\sum_J  X_J^{\rm d.ver}(\kappa_1,\kappa_2,\kappa_3)\,
  \nonumber \\ & \times
  \frac{
   R_J(\omega,an_3n_1a)\,R^{(1)}_{n_1n_2} \, R^{(-2)}_{n_2n_3}  }
 {(\vare-\omega-u\,\vare_{n_1})(\vare-\omega-u\,\vare_{n_2})(\vare-\omega-u\,\vare_{n_3})}
\,,
\end{align}
where
\begin{align}
& X_J^{\rm d.ver}(\kappa_1,\kappa_2,\kappa_3)
   = \frac{(-1)^{J+j_a-j_2}}{2}\,
 \nonumber \\ & \times
\sum_{j_n=1/2,3/2}(2j_n+1)
 \SixJ{1}{j_n}{j_a}{J}{j_1}{j_2}  \SixJ{1}{j_n}{j_a}{J}{j_3}{j_2}
 \nonumber \\ & \times
 \frac{\kappa_1+\kappa_2}{\sqrt{3}}\, C_1(-\kappa_2,\kappa_1)\,
 \frac{\kappa_2+\kappa_3}{\sqrt{3}}\, C_1(-\kappa_3,\kappa_2)\,.
\end{align}
Note that the expression for $\delta_1 \sigma$ is IR divergent when
any two or all three of the
intermediate states have the same energy as the reference state: $\vare_{n_1}
= \vare_{n_2} = \vare_a$, $\vare_{n_2} = \vare_{n_3} = \vare_a$, $\vare_{n_1}
= \vare_{n_3} = \vare_a$, $\vare_{n_1} = \vare_{n_2} = \vare_{n_3} =
\vare_a$. The IR divergence cancels when all parts of Eq.~(\ref{dvr}) are added together.

The contribution of the second term to the shielding is
\begin{align}    \label{delta2}
\delta_2\sigma &\ = \frac{\alpha}{2}\, \lbr \widetilde{V}_{\rm zee}\rbr\,
\lbr \widetilde{V}_{\rm hfs}\rbr\, \lbr \Sigma ^{\prime\prime}\rbr
= \frac{4}{9}\, R^{(1)}_{aa}\,R^{(-2)}_{aa}\,
 \nonumber \\ & \times
\frac{i\alpha^2}{2\pi} \int_{-\infty}^{\infty} d\omega
  \sum_{n_1n_2n_3}
\sum_J
\frac{(-1)^{J+j_a-j_1}}{2j_a+1}\,\delta_{\kappa_1,\kappa_2}\,\delta_{\kappa_1,\kappa_3}\,
 \nonumber \\ & \times
  \frac{
   R_J(\omega,an_3n_1a)\,N_{n_1n_2} \, N_{n_2n_3}  }
 {(\vare-\omega-u\,\vare_{n_1})(\vare-\omega-u\,\vare_{n_2})(\vare-\omega-u\,\vare_{n_3})}
\,,
\end{align}
where $N_{ab}$ is the normalization integral,
\begin{align}
N_{ab} = \int_0^{\infty}dx\, x^2\,(g_ag_b+f_af_b)\,.
\end{align}
The expression for $\delta_2\sigma$ is IR divergent when $\vare_{n_1} = \vare_{n_2} = \vare_{n_3} =
\vare_a$.

The third term is given by
\begin{align}    \label{delta3}
\delta_3 \sigma &\ = \frac13\, R_{aa}^{(1)}\,
\frac{i\alpha^2}{2\pi} \int_{-\infty}^{\infty} d\omega
  \sum_{n_1n_2n_3}
\nonumber \\ & \times
\biggl[
  \frac{\sum_J  X_{J}(\kappa_1,\kappa_2)\,\delta_{\kappa_2,\kappa_3}\,
   R_J(\omega,an_3n_1a)\,R^{(-2)}_{n_1n_2} \, N_{n_2n_3}  }
 {(\vare-\omega-u\,\vare_{n_1})(\vare-\omega-u\,\vare_{n_2})(\vare-\omega-u\,\vare_{n_3})}
\nonumber \\ &
+   \frac{\sum_J  X_{J}(\kappa_2,\kappa_3)\,\delta_{\kappa_1,\kappa_2}\,
   R_J(\omega,an_3n_1a)\, N_{n_1n_2}\,R^{(-2)}_{n_2n_3}   }
 {(\vare-\omega-u\,\vare_{n_1})(\vare-\omega-u\,\vare_{n_2})(\vare-\omega-u\,\vare_{n_3})}
\Biggr]
\,,
\end{align}
where
\begin{align}
X_J(\kappa_1,\kappa_2) =
\frac{1}{\sqrt{2}}
 \SixJ{j_1}{j_2}{1}{j_{a}}{j_a}{J}
\frac{-\kappa_1-\kappa_2}{\sqrt{3}} C_1(-\kappa_2,\kappa_1)\,.
\end{align}

The fourth term $\delta_4\sigma$ is obtained from $\delta_3\sigma$ by an
obvious substitution $R^{(1)}\leftrightarrow R^{(-2)}$.

The fifth term is given by
\begin{align}    \label{delta5}
\delta_5\sigma = -\sigma^{(0)}\, \frac{i\alpha}{2\pi}
\int_{-\infty}^{\infty} d\omega
  \frac{\sum_J  \frac{(-1)^{J}}{2j_a+1}\, R_J(\omega,aaaa)}{(-\omega+i0)^2}\,.
\end{align}

Finally, the total double-vertex contribution is given by the sum of the five
terms discussed above,
\begin{align}    \label{sum}
\Delta \sigma_{\rm d.vr} = \sum_{i = 1}^5\delta_i\sigma\,.
\end{align}
Despite the fact that the individual terms $\delta_i\sigma$ are IR divergent,
the sum of them is finite and can be evaluated without any explicit
regularization, provided that (i) the integration over the frequency of the
virtual photon in the self-energy loop $\omega$ is performed after all five
terms are added together and (ii) the contour of the $\omega$ integration is
suitably chosen. One can show that if the $\omega$ integration is performed
along the contour consisting of the low-energy and high-energy parts, as
e.g., in Refs.~\cite{yerokhin:05:se,yerokhin:05:hfs}, the integrand becomes regular
at $\omega\to 0$ and, therefore, can be directly evaluated numerically.

The numerical evaluation of the double-vertex correction $\Delta \sigma_{\rm
  d.vr}$ was the most time consuming part of our calculation, due to a large
number of the partial waves involved and a four-dimensional radial integration
in Eq.~(\ref{delta1}). The radial integration was carried out with help of the
numerical approach developed in our calculation of the two-loop self-energy
and described in detail in Ref.~\cite{yerokhin:03:epjd}. Because of
numerical cancellations between the five terms in Eq.~(\ref{sum}), especially
in the region of small values of $\omega$, we took
care to treat all five terms exactly in the same way. In particular, the
normalization integrals $N_{ab}$ in Eqs.~(\ref{delta2}) and (\ref{delta3})
were evaluated numerically, in order to be consistent with the evaluation
of Eq.~(\ref{delta1}),
whereas the corresponding contributions could have been evaluated more easily by taking
the derivative of the Dirac-Coulomb Green function.

\subsection{Vacuum polarization correction}

\begin{figure*}
\centerline{\includegraphics[width=0.8\textwidth]{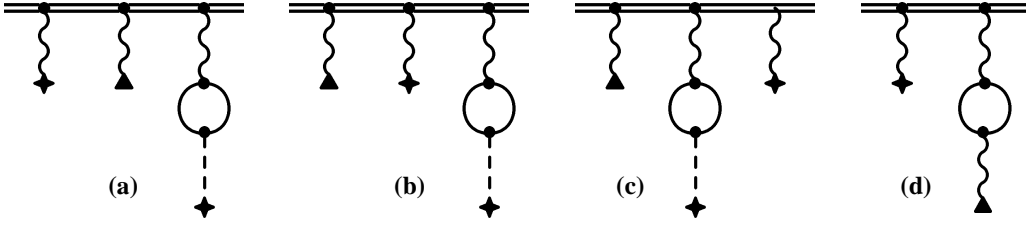}}
\caption{Vacuum-polarization corrections to the nuclear magnetic shielding
  calculated in the present work. The double line
  represents electron propagating in the binding nuclear field,
  the single line represents the free-electron propagator,
  the dashed line terminated by a cross represents interaction with the binding
  Coulomb field, the wave line terminated by a triangle represents the dipole
  hyperfine interaction with the nucleus, and
  the wave line terminated by a cross represents the interaction with the
  external magnetic field.
 \label{fig:2} }
\end{figure*}

Vacuum polarization corrections to the magnetic shielding calculated in the
present work are shown in Fig.~\ref{fig:2}. The diagrams in
Fig.~\ref{fig:2}(a)-(c) come from the insertion of the Uehling potential
into the electron line of the leading-order magnetic shielding, whereas the
diagram in Fig.~\ref{fig:2}(d) respresents the Uehling potential insertion
into the hfs interaction. We note that the diagram with the
vacuum polarization insertion into the Zeeman interaction
vanishes in the Uehling-potential approximation. In our treatment, we neglect
contributions with additional Coulomb interactions in the vacuum-polarization loop
(which correspond to the Wichmann-Kroll part of the one-loop vacuum
polarization) and the additional diagram of the Wichmann-Kroll type with both
hfs and Zeeman interactions attached to the vacuum-polarization loop.
We expect that the part accounted for in the present work yields the dominant
contribution to the vacuum polarization correction.

The contribution of the diagrams in Fig.~\ref{fig:2}~(a)-(c) is an analogue of
the perturbed-orbital self-energy contribution and is given by a similar
expression,
\begin{eqnarray}
\Delta \sigma_{\rm VP,po} &=& \alpha\, \lbr a_{\nicefrac12}|V_{\rm Ueh} |\delta^{(2)}a_{\nicefrac12}\rbr
 \nonumber \\ &&
 + \alpha\, \lbr \delta^{(1)}_{\rm hfs}a_{\nicefrac12}|V_{\rm Ueh} |\delta^{(1)}_{\rm zee}a_{\nicefrac12}\rbr\,,
\end{eqnarray}
where $V_{\rm Ueh}$ is the Uehling potential,
\begin{align}\label{c2}
   V_{\rm Ueh}(r) &\ = -\frac{2\alpha^2 Z}{3 m r}\,
    \int_0^{\infty} dr\pr r\pr \rho(r\pr)\,
  \nonumber \\ & \times
       \left[ K_0(2m|r-r\pr|)-K_0(2m|r+r\pr|)\right]\,,
\end{align}
and
\begin{equation}\label{c3}
 K_0(x) =
      \int_1^{\infty}dt\, e^{-xt}
         \left(\frac1{t^3}+\frac1{2t^5}\right)\,
     \sqrt{t^2-1}\,,
\end{equation}
and the nuclear-charge density $\rho$ is normalized by the condition $\int
d\vec{r}\,\rho(r) = 1$. We note that this contribution can also be
calculated by incorporating the Uehling potential into the Dirac equation and
re-evaluating the leading-order magnetic shielding, in this way accounting for
the Uehling potential to all orders. We performed calculations in both ways,
which ensured that the
perturbations of the reference-state wave function $|\delta^{(1)}a\rbr$ and
$|\delta^{(2)}a\rbr$ are computed correctly.

The contribution of the diagram in Fig.~\ref{fig:2}(d) to the magnetic
shielding can be expressed as
\begin{align}
\Delta \sigma_{\rm VP,mag} = &\,  \alpha
\sum_{n\ne a}\frac1{\vare_a-\vare_n}\,
 \nonumber \\ & \times
  \lbr a_{\nicefrac12}| \widetilde{V}_{\rm zee} |n_{\nicefrac12}\rbr
  \lbr n_{\nicefrac12}| \widetilde{V}_{\rm VP,mag} |a_{\nicefrac12}\rbr\,,
\end{align}
where
\begin{align}
\widetilde{V}_{\rm VP,mag}(r) = &\ \widetilde{V}_{\rm hfs}(r)\,\,
 \frac{2\alpha}{3\pi}\,\int_1^{\infty}dt\,
   \frac{\sqrt{t^2-1}}{t^2}\,
 \nonumber \\ & \times
\left( 1+ \frac1{2t^2}\right) \, (1+2mrt) \, e^{-2mrt}\,
\end{align}
is the hfs interaction modified by the vacuum-polarization insertion.

%
\section{QED correction for small nuclear charges}
\label{sec:Zaexp}

In the previous section, we calculated the QED corrections to the magnetic
shielding without any expansion in the nuclear binding strength parameter
$\Za$. Now we turn to the evaluation of this corrections within the expansion
in this parameter. We will derive the complete expression for the leading term
of the $\Za$ expansion, which enters in the relative order
$\delta \sigma/\sigma \sim \alpha(\Za)^2$. The results
obtained in this section
will be applicable for light hydrogen-like ions, where no all-order
calculations were possible because of large numerical cancellation. They will also
provide an important cross-check with the all-order calculations described in
the previous section.

For the derivation, it is convenient to use the formalism of the
non-relativistic quantum electrodynamics (NRQED). In this approach, all QED
effects are calculated by an expansion in powers of $\alpha$ and $\Za$
and represented as expectation values of various effective operators
on the nonrelativistic reference-state wave function.
Let us start with the effective Hamiltonian $H_{\rm NRQED}$, which
includes leading one-loop radiative corrections:
\begin{eqnarray}
H_{\rm NRQED} &=& \frac{\vec \pi^2}{2\,m}+e\,A^0
-\frac{e}{6}\,\biggl(\frac{3}{4\,m^2}+r_E^2\biggr)\,\vec\nabla\cdot\vec E
\nonumber \\ &&
+\frac{e}{12\,m}\biggl(r_E^2-\frac{3\,\kappa}{4\,m^2}\biggr)
\bigl\{\vec \pi\,,\,\vec\nabla\times\vec B\bigr\}
\nonumber \\ &&
-\frac{e^2}{2}\,\biggl(\frac{1}{4\,m^3}+\alpha_M\biggr)\,\vec B^2\,,
\label{t02}
\end{eqnarray}
where $\{\,.\,,.\}$ denotes anticommutator,
$\vec \pi = \vec p - e\,\vec A$. The QED effects in the above Hamiltonian are
parameterized in terms of the constants $\kappa$, $r_E$, and $\alpha_M$, which
are interpreted as the electron magnetic moment anomaly, the charge radius and the
magnetic polarizability, respectively,
\begin{eqnarray}
\kappa &=& \frac{\alpha}{2\,\pi}\,,\\
r_E^2 &=& \frac{3\,\kappa}{2\,m^2} + \frac{2\,\alpha}{\pi\,m^2}\,
\biggl(\ln\frac{m}{2\,\epsilon} +\frac{11}{24}\biggr)\,,\label{t03}\\
\alpha_M &=& \frac{4\,\alpha}{3\,\pi\,m^3}\,
\biggl(\ln\frac{m}{2\,\epsilon} +\frac{13}{24}\biggr)\,,
\end{eqnarray}
where $\epsilon$ is the photon momentum cutoff. In the non-QED limit,
all QED constants vanish,
$r^2_E=\alpha_M=\kappa=0$, and the effective Hamiltonian
$H_{\rm NRQED}$ turns into the Schr{\"o}dinger-Pauli Hamiltonian.

The QED constants $r_E^2$ and $\alpha_M$ depend explicitly on the
the photon momentum cutoff parameter $\epsilon$. This dependence cancels
out with contributions coming from the emission and absorption of
virtual photons with the energy smaller than $\epsilon$.
The complete expression for any physical quantity,
e.g., the Lamb shift and the shielding constant, does not depend on the
artificial photon cutoff parameter.

The above formula for $r^2_E$ is obtained from the well known expressions
for the one-loop radiative corrections
to electromagnetic formfactors $F_1$ and $F_2$ \cite{itzykson:80}.
The formula for $\alpha_M$ has not appeared in the literature previously.
We have derived it by a method similar to that used
in Ref.~\cite{jentschura:05:sese} for the electric polarizability, denoted by $\chi$ in that work.

We will start with rederiving the known nonrelativistic expression for the
shielding constant within our approach.
It comes from the $\vec A^2$ term in the electron kinetic energy
in Eq. (\ref{t02}).
The electromagnetic potential $\vec A$ is the sum of
the external magnetic potential $\vec A_E$,
\begin{equation}
\vec A_E = \frac{1}{2}\,\vec B\times\vec r\,,
\end{equation}
and the potential induced by the nuclear magnetic moment,
\begin{equation}
\vec A_I = \frac{1}{4\,\pi}\,\vec\mu\times\frac{\vec r}{r^3}\,.
\end{equation}
The corresponding energy shift is
\begin{align}
\delta E &\ =
\frac{e^2}{2\,m}\langle {\vec A\,}^2\rangle
=\frac{e^2}{m}\langle \vec A_E \cdot \vec A_I\rangle
 \nonumber \\ &
=\frac{\alpha}{2\,m}\,\biggl\langle \left(\vec B\times\vec r\right)
\cdot \left(\vec\mu\times\frac{\vec r}{r^3}\right)\biggr\rangle,
\end{align}
where matrix elements are calculated with the nonrelativistic wave function.
The shielding constant $\sigma$ is obtained from the energy shift $\delta E$ by
$\delta E = \sigma\,\vec\mu\cdot\vec B$.
For the ground $(L=0)$ hydrogenic state, the nonrelativistic results is
\begin{equation}
\sigma = \frac{\alpha}{3\,m}\,\biggl\langle
\frac{1}{r}\biggr\rangle.
\end{equation}

Before considering the QED effects to the shielding constant,
it is convenient to first recalculate the leading QED correction
to the energy levels. The total contribution is split into two parts induced
by the virtual photons of low ($L$) and high ($H$) energy,
\begin{equation}
\delta E_{\rm Lamb} =  \delta E_L + \delta E_H\,.
\end{equation}
The high-energy part is the expectation value of the
$\vec\nabla\cdot\vec E$ term in
the effective Hamiltonian in Eq. (\ref{t02})
\begin{eqnarray}
\delta E_H &=& \biggl\langle
-\frac{e}{6}\,r_E^2\,\vec\nabla\cdot\vec E\biggr\rangle
=\frac{2}{3}\,\frac{(Z\,\alpha)^4}{n^3}\,r_E^2\,\delta_{l0}
\nonumber \\ &&
=\frac{\alpha}{\pi}\,\frac{(Z\,\alpha)^4}{n^3}\,
\biggl(\frac{4}{3}\,\ln\frac{m}{2\,\epsilon}+\frac{10}{9}\biggr)\,\delta_{l0}\,,
\end{eqnarray}
where $\vec E=-\vec \nabla A^0$ and $A^0 = -Z\,e/(4\,\pi\,r)$.
The vacuum polarization can be incorporated in the above expression
by adding $-2\,\alpha/(5\,\pi\,m^2)$ to $r^2_E$ in Eq. (\ref{t03}).
The low-energy part $\delta E_L$ is induced by the emission and the
absorption of the virtual photons of low $(k<\epsilon)$ energy,
\begin{align}
&\delta E_L = e^2\int_0^\epsilon
\frac{d^3k}{(2\,\pi)^3\,2\,k}\,
\biggl(\delta^{ij}-\frac{k^i\,k^j}{k^2}\biggr)\,
\biggl\langle\frac{p^i}{m}\,
\frac{1}{E-H-k}
\,\frac{p^j}{m}\biggr\rangle
  \nonumber \\ &
=
\frac{2\,\alpha}{3\,\pi}\,
\biggl\langle\frac{\vec p}{m}\,(H-E)\,\biggl\{
\ln\biggl[\frac{2\,\epsilon}{m\,(Z\,\alpha)^2}\biggr]
-\ln\biggl[\frac{2\,(H-E)}{m\,(Z\,\alpha)^2}\biggr]\biggr\}
\frac{\vec p}{m}\biggr\rangle\,.
\end{align}
The terms containing $\ln\epsilon$ cancel out in the sum of the low- and
high-energy parts, as expected.
The total leading-order Lamb shift contribution is
\begin{eqnarray}
\delta E_{\rm Lamb} &=& \frac{\alpha}{\pi}\,\frac{(Z\,\alpha)^4}{n^3}\,
\biggl\{\frac{4}{3}\ln\bigl[(Z\,\alpha)^{-2}\bigr]\,\delta_{l0}
\nonumber \\&&
+\biggl(\frac{10}{9}-\frac{4}{15}\biggr)\,\delta_{l0}
-\frac{4}{3}\,\ln k_0(n,l)\biggr\}\,,
\end{eqnarray}
where $\ln k_0(n,l)$ is the Bethe logarithm,
\begin{equation}
\ln k_0(n,l) = \frac{n^3}{2\,m^3\,(Z\,\alpha)^4}\,
\Bigl\langle\vec p\,(H-E)
\ln\biggl[\frac{2\,(H-E)}{m\,(Z\,\alpha)^2}\biggr]
\vec p\,\Bigr\rangle\,.
\end{equation}

We now turn to the calculation of the
QED correction to the magnetic shielding, which is performed
similarly to that for the Lamb shift. The total contribution is
split into the low and the high-energy parts:
\begin{equation}
\delta E = \delta E_L +\delta E_H\,,
\end{equation}
where
\begin{align}
\delta E_L &\ = e^2\,\int_0^\epsilon \frac{d^3 k}{(2\,\pi)^3\,2\,k}\,
\biggl(\delta^{ij}-\frac{k^i\,k^j}{k^2}\biggr)\,
\biggl\langle\frac{\pi^i}{m}\,
\frac{1}{E-H-k}
\,\frac{\pi^j}{m}\biggr\rangle\,,
\end{align}
and
\begin{align}
\delta E_H &\ = -2\, \biggl\langle \frac{e}{6}\,r^2_E\,\vec\nabla\cdot\vec E\,
\frac{1}{(E-H)'}\,\frac{{\vec A\,}^2}{2\,m}\biggr\rangle
\nonumber \\ &
+\biggl\langle\frac{e}{12\,m}\,\biggl(
r^2_E-\frac{3\,\kappa}{4\,m^2}\biggr)\,
\bigl\{\vec\pi\,,\,\vec\nabla\times\vec B\bigr\}\biggr\rangle
-\biggl\langle
\frac{e^2}{2}\,\alpha_M\,{\vec B\,}^2\biggr\rangle \,.
\end{align}

The high-energy part can be conveniently rewritten in the form
\begin{eqnarray}
\delta E_H &=& \frac{1}{9}\,Z\,\alpha^2\,r^2_E\,
\biggl\langle\frac{1}{r}\,\frac{1}{(E-H)'}\,4\,\pi\,\delta^3(r)\biggr\rangle
\,\vec\mu\cdot\vec B \nonumber \\
&&-\frac{2\,\pi\,\alpha}{3\,m}\,\biggl(r^2_E-\frac{3\,\kappa}{4\,m^2}\biggr)
\,\bigl\langle\vec A_E\cdot\vec\nabla\times\vec B_I\bigr\rangle
\nonumber \\&&
-\,4\,\pi\,\alpha\,\alpha_M\,\bigl\langle\vec B\cdot\vec B_I\bigr\rangle  \,.
\end{eqnarray}

In the low-energy part, we separate out $\ln\epsilon$
and then perform an expansion in the magnetic fields,
\begin{eqnarray}
\delta E_L &=& \delta E_{LA} + \delta E_{LB},\\
\delta E_{LA} &=& -\frac{2\,\alpha}{3\,\pi}\,
\biggl\langle\frac{\vec\pi}{m}\,(H-E)\,\ln\frac{2\,(H-E)}{m\,(Z\,\alpha)^2}\,
\frac{\vec\pi}{m}\biggr\rangle,\\
\delta E_{LB} &=&\frac{2\,\alpha}{3\,\pi}\,
\biggl\langle\frac{\vec\pi}{m}\,(H-E)\,
\frac{\vec\pi}{m}\biggr\rangle\,
\ln\frac{2\,\epsilon}{m\,(Z\,\alpha)^2}\,.
\end{eqnarray}
Using the identity
\begin{align}
&2\,\biggl\langle\frac{\vec\pi}{m}\,(H-E)\,\frac{\vec\pi}{m}\biggr\rangle
= \biggl\langle\biggl[\frac{\vec\pi}{m}\,,
\biggl[H-E\,,\,\frac{\vec\pi}{m}\biggr]\biggr]\biggr\rangle
\nonumber \\
&=\biggl\langle4\,\pi\,Z\,\alpha\,\delta^3(r)
+ \frac{e}{2\,m^3}\Bigl\{\vec \pi\,,\,\vec\nabla\times\vec B\Bigr\}
+ \frac{2\,e^2}{m^3}\,B^2 \biggr\rangle\,,
\end{align}
we transform $\delta E_{LB}$ to the form
\begin{eqnarray}
\delta E_{LB} &=& \frac{\alpha}{3\,\pi}\,
\ln\biggl[\frac{2\,\epsilon}{m\,(Z\,\alpha)^2}\biggr]
\nonumber \\&&
\biggl[\,2\,\biggl\langle
\frac{\alpha}{3\,r}\,\frac{1}{(E-H)'}
\,4\,\pi\,Z\,\alpha\,\delta^3(r)\biggr\rangle\,\vec\mu\cdot\vec B\\\nonumber
&&-4\,\pi\,\alpha\,\bigl\langle\vec A_E\cdot\vec\nabla\times\vec B_I\bigr\rangle
+16\,\pi\,\alpha\,\bigl\langle\vec B\cdot\vec B_I\bigr\rangle\biggr]\,.
\end{eqnarray}

The artificial parameter $\ln\epsilon$ cancels out in the sum $\delta E_{LB}+
\delta E_{H}$ separately for each type of the matrix elements,
\begin{align}
&\delta E_{LB}+ \delta E_{H} = \frac{2}{9}\,\frac{\alpha^2}{\pi}\,Z\,\alpha\,
\nonumber \\ &\times
\biggl[\ln(Z\,\alpha)^{-2}+\frac{5}{6}-\frac{1}{5}\biggr]\,
\vec\mu\cdot\vec B\,
\biggl\langle 4\,\pi\,\delta^3(r)\,\frac{1}{E-H)'}\,\frac{1}{r}\biggr\rangle \nonumber \\
& -\frac{4}{3}\,\alpha^2\,\biggl[
\ln(Z\,\alpha)^{-2}+\frac{31}{48}-\frac{1}{5}\biggr]\,
\bigl\langle\vec A_E\cdot\vec\nabla\times\vec B_I\bigr\rangle \nonumber \\
& +\frac{16}{3}\,\alpha^2\,
\biggl[\ln(Z\,\alpha)^{-2}-\frac{13}{24}\biggr]\,\bigl\langle\vec B\cdot\vec B_I\bigr\rangle.
\end{align}
Using the following results for the matrix elements with $nS$ states
\begin{eqnarray}
\biggl\langle 4\,\pi\,\delta^3(r)\,\frac{1}{E-H)'}\,\frac{1}{r}\biggr\rangle &=&
-\frac{6\,(Z\,\alpha)^2}{n^3},\\
\bigl\langle\vec A_E\cdot\vec\nabla\times\vec B_I\bigr\rangle &=&
\frac{(Z\,\alpha)^3}{\pi\,n^3}\,\vec\mu\cdot\vec B,\\
\bigl\langle\vec B\cdot\vec B_I\bigr\rangle &=& \frac{2}{3\,\pi}\,
\frac{(Z\,\alpha)^3}{n^3}\,\vec\mu\cdot\vec B\,,
\end{eqnarray}
we obtain
\begin{equation}
\delta E_{LB}+\delta E_H  = \frac{8\,\alpha^2}{9\,\pi}\,
\frac{(Z\,\alpha)^3}{n^3}\,
\biggl[ \ln(Z\,\alpha)^{-2}-\frac{421}{96}+\frac{3}{5}\biggr]\,\vec\mu\cdot\vec B\,.
\end{equation}

The calculation of the remaining low-energy contribution $\delta E_{LA}$ is
slightly more complicated.
We first return to the integral form of $\delta E_L$, derive an expression for
$\delta E_{LA}$, and then drop all terms with $\ln\epsilon$,
as they are already accounted for by $\delta E_{LB}$. The integral form of
$\delta E_L$ is
\begin{equation}
\delta E_L =
\frac{2\,\alpha}{3\,\pi}\,\int_0^\epsilon k\,dk\,
\biggl\langle\frac{\vec\pi}{m}\,\frac{1}{E-H-k}\,\frac{\vec \pi}{m}\biggr\rangle.
\end{equation}
It can be rewritten, using the identity $\vec\pi = -i\,m\,[\vec r\,,\,H-E]$,
in the form
\begin{equation}
\delta E_L =
\frac{2\,\alpha}{3\,\pi}\,\int_0^\epsilon k^3\,dk\,
\biggl\langle\vec r\,\frac{1}{E-H-k}\,\vec r\biggr\rangle. \label{t34}
\end{equation}
All terms with positive powers of $\epsilon$ are discarded,
since one assumes that the limit $\epsilon\rightarrow 0$ is taken
after the expansion in $\alpha$ is done.
We shall now expand the integrand in Eq. (\ref{t34}) in the magnetic
fields. The Hamiltonian $H$ is
\begin{equation}
H = H_0 +\frac{\alpha}{3\,r}\,\vec\mu\cdot\vec B
-\frac{e}{2\,m}\,\vec L\cdot\vec B
-\frac{e}{4\,\pi\,m\,r^3}\,\vec L\cdot\vec\mu \label{t35}\,,
\end{equation}
where $H_0 = p^2/(2\,m)-Z\,\alpha/r$.
The first term with $\vec B$ in Eq. (\ref{t35})
can be absorbed into $Z' = Z-\vec\mu\cdot\vec B/3$.
So, the correction due to this term is
\begin{eqnarray}
\delta E_{L1} &=&
\frac{4\,\alpha}{3\,\pi}\,(Z'\,\alpha)^4\,
\biggl\{\ln\biggl[\frac{2\,\epsilon}{m\,(Z'\,\alpha)^2}\biggr]-\ln k_0\biggr\}
\\ &=&
\frac{16\,\alpha^2}{9\,\pi}\,\frac{(Z\,\alpha)^3}{n^3}\,\vec\mu\cdot\vec B\,
\biggl\{\ln k_0 +\frac{1}{2}-\ln\biggl[\frac{2\,\epsilon}{m\,(Z\,\alpha)^2}\biggr]\biggr\}\,.
\nonumber
\end{eqnarray}
The correction due to the two last terms in Eq. (\ref{t35}) is
\begin{widetext}
\begin{align}
\delta E_{L2} &\ =
\frac{2\,\alpha^2}{3\,\pi\,m^2}\,
\int_0^\epsilon dk\,k^3\,\biggl\langle\vec r
\frac{1}{E-H-k}\,\vec L\cdot\vec B
\frac{1}{E-H-k}\,
\frac{\vec\mu\cdot\vec L}{r^3}\,\frac{1}{E-H-k}\,\vec r\biggr\rangle
   \nonumber\\
&\ =\frac{2\,\alpha^2}{9\,\pi}\,\vec\mu\cdot\vec B\,
\int_0^\epsilon dk\,k^3
\frac{d}{dk}\,\biggl\langle\vec r\,
\frac{1}{E-H-k}\,\frac{1}{r^3}\,\frac{1}{E-H-k}\,\vec r\biggr\rangle\,.
\end{align}
\end{widetext}
Integrating by parts and using the results and
the notation of $\ln k_3$ from Ref. \cite{pachucki:05:gfact}
\begin{align}
&\int_0^\epsilon dk\,k^2 \biggl\langle\vec r\,
\frac{1}{E-H-k}\,\frac{1}{r^3}\,\frac{1}{E-H-k}\,\vec r\biggr\rangle
\nonumber \\ &=
\epsilon\,\biggl\langle\frac{1}{r}\biggr\rangle-\frac{4\,(Z\,\alpha)^3}{n^3}\,
\biggl\{\ln\biggl[\frac{2\,\epsilon}{m\,(Z\,\alpha)^2}\biggr]-\ln k_3\biggr\}\,,
\end{align}
we obtain
\begin{equation}
\delta E_{L2} = \frac{8\,\alpha^2}{3\,\pi\,m^2}\,\frac{(Z\,\alpha)^3}{n^3}\,
\vec\mu\cdot\vec B\,\biggl\{
\ln\biggl[\frac{2\,\epsilon}{m\,(Z\,\alpha)^2}\biggr]-\ln k_3-\frac{1}{3}\biggr\}\,.
\end{equation}
The sum of these two contributions $\delta E_{LA} = \delta E_{L1}+\delta
E_{L2}$, after dropping the $\ln\epsilon$ terms, is
\begin{equation}
\delta E_{LA} =\frac{8\,\alpha^2}{9\,\pi}\,\frac{(Z\,\alpha)^3}{n^3}\,
\vec\mu\cdot\vec B\,\bigl(2\,\ln k_0-3\,\ln k_3\bigr).
\end{equation}
Finally, the total correction to energy is
\begin{equation}
\delta E = \delta E_{LA} + \delta E_{LB} +\delta E_{H}\,.
\end{equation}
Expressing the energy shift in terms of the shielding constant ($\delta E =
\delta\sigma\,\vec\mu\cdot\vec B$), we obtain the complete result for the
leading QED correction to the nuclear magnetic shielding in hydrogen-like ions,
valid for the $nS$ states,
\begin{align}
\delta\sigma &\ =
\frac{8\,\alpha^2}{9\,\pi}\,\frac{(Z\,\alpha)^3}{n^3} \label{t42}
\\ & \times
\biggl\{
\ln\bigl[(Z\,\alpha)^{-2}\bigr]+2\,\ln k_0(n)-3\,\ln k_3(n)-\frac{421}{96}+\frac{3}{5}\biggr\}\,.
\label{eq:Za}
\end{align}
The term of $3/5$ in the brackets is the contribution of the vacuum
polarization.
The numerical results for the Bethe logarithm $\ln k_0$ and the $1/r^3$
Bethe-logarithm-type correction $\ln k_3$ \cite{pachucki:05:gfact}  for the $1s$ state are
 \begin{eqnarray}
\ln k_0(1s) &=&  2.984\,128\,556,\\
\ln k_3(1s) &=& 3.272\,806\,545\,.
\end{eqnarray}
We note that the numerical value of the constant term in Eq.~(\ref{eq:Za}),
$-7.635\,58$, is comparable in magnitude but of the opposite sign to
the logarithmic term at $Z=1$, $\ln\alpha^{-2} = 9.840\,49$. This entails
a significant numerical cancellation between these two terms
for hydrogen and light hydrogen-like ions. As a result, the total QED
correction turns out to be much smaller in magnitude than could be anticipated
from the leading logarithm alone.

\section{Other corrections}
\label{sec:3}

\subsection{Bohr-Weisskopf correction}

We now turn to the effect induced by the spatial distribution of the
nuclear magnetic moment, also known as the Bohr-Weisskopf (BW)
correction. Our treatment of the BW effect is based on the effective
single-particle (SP) model of the nuclear magnetic moment.
Within this model, the
magnetic moment is assumed to be induced by the odd nucleon (proton, when
$Z$ and $A$ are odd and neutron, when $Z$ is even and $A$ is odd) with
an effective $g$-factor, which is adjusted to yield the
experimental value of the nuclear magnetic moment.
The treatment of the
magnetization distribution effect on hfs within the SP model was originally
developed in Refs.~\cite{bohr:50,bohr:51} and later in Ref.~\cite{shabaev:94:hfs}.
Our present treatment closely follows the procedure
described in Refs.~\cite{shabaev:94:hfs,shabaev:97:pra,zherebtsov:00:cjp}.

The wave function of the odd nucleon is assumed to satisfy the Schr\"odinger
equation with the central potential of Woods-Saxon form and the spin-orbital
term included (see, e.g., Ref.~\cite{elton:67})
\begin{equation}\label{5eq10}
    V(\bfr) = -V_0\,{\cal F}(r)+ \frac1{m_p}\phi_{so}(r)\,\vec{l}\cdot\bsigma+ V_C(r)\,,
\end{equation}
where
\begin{equation}\label{5eq10a}
    \phi_{so}(r) = \frac{V_{so}}{4m_pr}\,
    \frac{d{\cal F}(r)}{dr}\,,
\end{equation}
\begin{equation}\label{5eq10b}
    {\cal F}(r) = \left[1+\exp\left(\frac{r-R}{a} \right)\right]^{-1}\,,
\end{equation}
and $V_C$ is the Coulomb part of the interaction (absent for neutron), with the
uniform distribution of the charge ($Z-1$) over the nuclear sphere. The parameters
$V_0$, $V_{so}$, $R$, and $a$ were taken from Refs.~\cite{elton:67,rost:68}.

The nuclear magnetic moment can be evaluated within the SP model to yield
\cite{shabaev:97:pra}
\begin{align}\label{5eq11}
    \frac{\mu}{\mu_N} = \left\{
     \begin{aligned}
      \displaystyle
       \frac12\, g_S+ \left[I-\frac12+\frac{2I+1}{4(I+1)}\lbr\phi_{so}r^2\rbr
       \right] g_L\,, \ \ \ \ \ \ \ \ \ \ \ & \\
        \ \mbox{for} \  \ I = L+\frac12\,, & \\
      \displaystyle
       -\frac{I}{2(I+1)}\, g_S+ \left[\frac{I(2I+3)}{2(I+1)}
        -\frac{2I+1}{4(I+1)}\lbr\phi_{so}r^2\rbr
       \right]& \ g_L\,,\ \\
       \ \mbox{for} \  \ I = L-\frac12\,, & \\
     \end{aligned}
\right.
\end{align}
where $I$ and $L$ are the total and the orbital angular momentum of the nucleus,
respectively, $g_L$ is the $g$ factor associated with the orbital motion of the
nucleon ($g_L$ = 1 for proton and $g_L = 0$ for neutron) and $g_S$ is the
effective nucleon $g$ factor, determined by the condition that Eq.~(\ref{5eq11})
yields the experimental value of the magnetic moment.

It was demonstrated in Ref.~\cite{shabaev:97:pra} that, within the SP model, the
BW effect can be accounted for by adding a multiplicative
magnetization-distribution function $F(r)$ to the standard point-dipole hfs
interaction. The distribution function is given by
\cite{zherebtsov:00:cjp}
\begin{widetext}
\begin{eqnarray} \label{5eq12}
  F(r) &=& \frac{\mu_N}{\mu}
    \int_0^{r}dr\pr\, {r\pr}^2 |u(r\pr)|^2\,
         \left[ \frac12\, g_S+ \left(I-\frac12+\frac{2I+1}{4(I+1)}\,r^2\phi_{so}(r)
       \right) g_L \right]
  \nonumber \\ &&
  + \frac{\mu_N}{\mu} \int_r^{\infty}dr\pr\, {r\pr}^2 |u(r\pr)|^2\,\frac{{r}^3}{{r\pr}^3}\,
         \left[ -\frac{2I-1}{8(I+1)}\, g_S+ \left(I-\frac12+\frac{2I+1}{4(I+1)}\,r^2\phi_{so}(r)
       \right) g_L \right]\,,
\end{eqnarray}
for $I = L+1/2$ and
\begin{eqnarray} \label{5eq13}
  F(r) &=& \frac{\mu_N}{\mu}
    \int_0^{r}dr\pr\, {r\pr}^2 |u(r\pr)|^2\,
         \left[ -\frac{I}{2(I+1)}\, g_S+ \left(\frac{I(2I+3)}{2(I+1)}-\frac{2I+1}{4(I+1)}\,r^2\phi_{so}(r)
       \right) g_L \right]
  \nonumber \\ &&
  + \frac{\mu_N}{\mu} \int_r^{\infty}dr\pr\, {r\pr}^2 |u(r\pr)|^2\,\frac{{r}^3}{{r\pr}^3}\,
         \left[ \frac{2I+3}{8(I+1)}\, g_S+ \left(\frac{I(2I+3)}{2(I+1)}-\frac{2I+1}{4(I+1)}\,r^2\phi_{so}(r)
       \right) g_L \right]\,,
\end{eqnarray}
\end{widetext}
for $I = L-1/2$. In the above formulas, $u(r)$ is the wave function of the odd
nucleon. It can easily be seen that $F(r) = 1$ outside the nucleus.

\subsection{Recoil and quadrupole corrections}

The recoil correction to the magnetic shielding was obtained in Ref.~\cite{rudzinski:09}
in the nonrelativistic approximation,
\begin{eqnarray} \label{eqrec}
\delta \sigma_{\rm rec} = -\frac{\alpha \Za}{3}\,\frac{m}{M}\,
\left( 1+ \frac{g_N-1}{g_N}\right)\,,
\end{eqnarray}
where $M$ is the nuclear mass and
\begin{eqnarray}
g_N = \frac{M}{Zm_p}\,\frac{\mu}{\mu_N \,I}\,.
\end{eqnarray}

The electric-quadrupole correction to the magnetic shielding is
\begin{eqnarray} \label{eqQ}
\delta \sigma_Q = -\frac{3Q}{2I(2I-1)}\,\frac{\delta g_Q}{(m/m_p)g_I}\,,
\end{eqnarray}
where $Q$ is the quadrupole moment of the nucleus and
$\delta g_Q$ is the quadrupole correction to the $g$ factor calculated
in Ref.~\cite{moskovkin:04},
\begin{align}
\delta g_Q &\ = \alpha\,(\Za)^3 \,\frac{12 \left[35+20\gamma-32(\Za)^2\right]}
 {135\,\gamma (1+\gamma)^2\left[15-16(\Za)^2\right]}
 \nonumber \\ &
 = \alpha\,(\Za)^3 \,\left[\frac{11}{135}+ \frac{43}{405}(\Za)^2+\ldots\right]\,.
\end{align}

\section{Results and discussion}
\label{sec:4}

Numerical results for the self-energy correction to the nuclear magnetic
shielding can be conveniently
parameterized in terms of the dimensionless function $D_{\rm SE}(\Za)$ defined
as
\begin{align}     \label{eqDZa}
\Delta \sigma_{\rm SE} = \alpha^2\,(\Za)^3\,D_{\rm SE}(\Za)\,.
\end{align}
Our numerical all-order (in $\Za$)
results for the self-energy correction to the magnetic
shielding are summarized in
Table~\ref{tab:se}. We note significant numerical cancellation between the
individual contributions, which is particularly strong for low values of $Z$. The fact that
the resulting sum is a smooth function of $Z$ and demonstrates the expected
$Z$ scaling serves as a
consistency check of our calculations. Besides of numerical cancellation, the
additional complication arising in the low-$Z$ region is that the
convergence of the partial-wave expansions becomes slower when $Z$
decreases. Because of these two complications, the accuracy of our
results worsens for smaller
values of $Z$ and no all-order results were obtained for $Z<10$.

\begin{figure}[t]
\centerline{\includegraphics[width=\columnwidth]{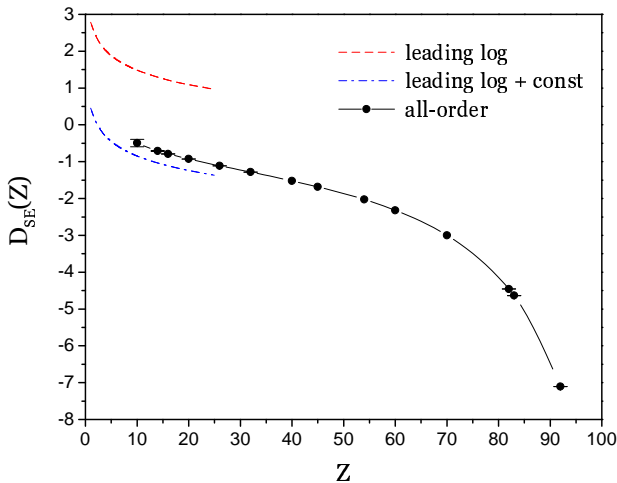}}
\caption{(Color online) Self-energy correction to the nuclear magnetic
  shielding.
 \label{fig:se} }
\end{figure}

The all-order numerical results can be compared with the
$\Za$-expansion results obtained in Sec.~\ref{sec:Zaexp}. As follows from
Eq.~(\ref{eq:Za}), the
$\Za$ expansion of the function $D_{\rm SE}$ for the $1s$ state reads
\begin{align}    \label{seZa}
D_{\rm SE}(\Za) = &\ \frac{8}{9\pi}\, \biggl[ \ln(\Za)^{-2}
-8.235\,579
+O(\Za) \biggr]\,,
\end{align}
where $O(\Za)$ denotes the higher-order terms.
In Fig.~\ref{fig:se}, the numerical all-order results for the function $D_{\rm SE}$
are plotted together with the contribution of the leading logarithmic
term in Eq.~(\ref{seZa}) (dashed line, red) and the contribution of both terms
in Eq.~(\ref{seZa}) (dashed-dot line, blue). We observe that the leading
logarithm alone gives a large contribution that disagrees strongly with the
all-order results. However, when the constant term is added,
the total $\Za$-expansion contribution shrinks significantly
and even changes its sign for $Z>3$. Only after the constant term
is accounted for, we observe reasonable agreement between
the all-order and $\Za$ expansion results.

%
%

%
%
\begin{figure}[t]
\centerline{\includegraphics[width=\columnwidth]{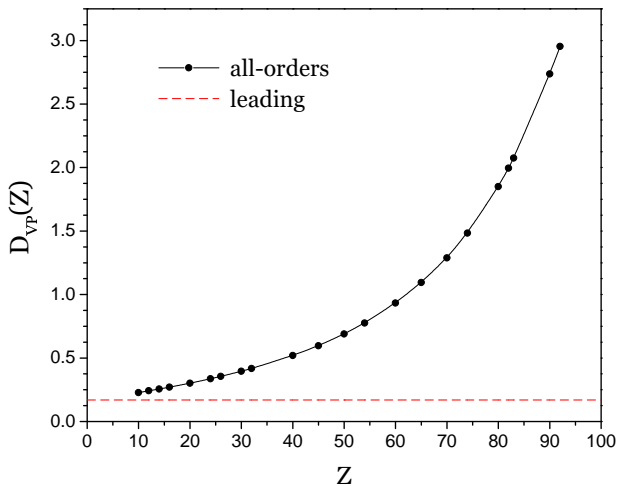}}
\caption{(Color online) Vacuum-polarization correction to the nuclear
magnetic shielding.
 \label{fig:vp} }
\end{figure}

\begin{figure}
\centerline{\includegraphics[width=\columnwidth]{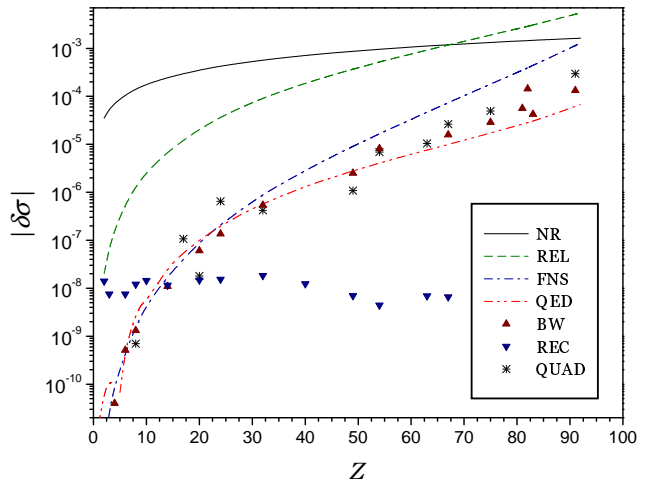}}
\caption{(Color online) Individual contributions to the nuclear shielding.
``NR'' is the nonrelativistic contribution,
  ``REL'' is the relativistic point-nucleus contribution, ``FNS'' is the
  finite nuclear size correction, ``QED'' is the QED correction, ``BW'' is the
  Bohr-Weisskopf correction, ``REC'' is the recoil correction, and ``QUAD'' is the
  electric quadrupole correction. Note that the QED correction changes its
  sign between $Z=4$ and 5.
 \label{fig:total} }
\end{figure}

The vacuum-polarization correction to the nuclear magnetic shielding is
parameterized as
\begin{align}  \label{eq:vp}
\Delta \sigma_{\rm VP} = \alpha^2\,(\Za)^3\,D_{\rm VP}(\Za)\,.
\end{align}
Our numerical results for the vacuum-polarization correction are presented in
Table~\ref{tab:vp}. The calculation was performed for the extended nucleus and
the Uehling potential was included to all orders. Note that our present treatment is
not complete, as the Wichmann-Kroll part of the correction is still
missing. We estimate the uncertainty due to omitted terms to be within 30\% of
the calculated contribution.

The $\Za$ expansion of the function $D_{\rm VP}$ reads
\begin{align}
D_{\rm VP}(\Za) = \frac{8}{9\pi n^3}\, \frac35\, \delta_{l,0} + O(\Za)\,,
\end{align}
where the Kronecker symbol $\delta_{l,0}$ indicates that the correction
vanishes for the reference states with $l>0$.
Comparison of the all-order numerical results
with the leading term of the $\Za$ expansion is
given in Fig.~\ref{fig:vp}. It is remarkable that
the all-order results grow fast when $Z$ is increased and eventually
become more than an order of magnitude larger than the leading-order result.
In the high-$Z$ region, the self-energy and vacuum-polarization
corrections are (as usual) of the opposite sign, the self energy being about twice
larger than the vacuum polarization.
In the low-$Z$ region, however, the self-energy correction changes its sign
and the total QED contribution becomes positive for very light ions.

The summary of our calculations of the nuclear magnetic shielding constant
$\sigma$ for several hydrogen-like ions is given in
Table~\ref{tab:total}. The first line of the table presents results for the
leading-order nuclear shielding, including the finite nuclear size effect. The
leading-order
contribution was calculated using the Fermi model for the nuclear charge
distribution and the point-dipole approximation for the interaction with the
nuclear magnetic moment. The nuclear charge radii were taken from
Ref.~\cite{angeli:04}. The results are in good agreement with those reported
in Ref.~\cite{moskovkin:04}. The QED correction presented in the second line
of Table~\ref{tab:total} is the sum of the all-order results for the
self energy and the vacuum polarization. Its error comes from the
numerical uncertainty of the self-energy part and the estimate of uncalculated
vacuum-polarization diagrams. Since there were no all-order calculations
performed for
oxygen, we used an extrapolation of our all-order results
(taking into account the derived values of the $\Za$ expansion coefficients).

The Bohr-Weisskopf correction, presented in the third line of
Table~\ref{tab:total}, was calculated by reevaluating the leading-order
contribution with the point-dipole hfs interaction modified by the
extended-distribution function $F(r)$ given by Eqs.~(\ref{5eq12}) and
(\ref{5eq13}). Because the effective single-particle model of the nuclear
magnetic moment is (of course) a rather crude approximation, we estimate the
uncertainty of the Bohr-Weisskopf correction to be 30\%, which is consistent
with previous estimates of the uncertainty of
this effect \cite{shabaev:97:pra}. This uncertainty includes also the error
due to the nuclear polarization effect, which is not considered in the present work.

The quadrupole and the recoil corrections, given in the fourth and fifth lines
of Table~\ref{tab:total}, respectively, were evaluated according to
Eqs.~(\ref{eqQ}) and (\ref{eqrec}). The error of the quadrupole
correction comes from the nuclear quadrupole moments. The magnetic dipole
and electric quadrupole moments of nuclei were taken from
Refs.~\cite{raghavan:89,stone:05}.

The $Z$ dependence of individual contributions to the magnetic
shielding constant $\sigma$ is shown in Fig.~\ref{fig:total}. The
leading-order contribution is separated into three parts, the point-nucleus
nonrelativistic part (solid line, black), the point-nucleus relativistic part
(dashed line, green), and the finite nuclear size correction (dashed-dotted
line, blue). We observe that the finite nuclear size correction, as well as
the other effects calculated in this work, become increasingly important in
the region of large nuclear charges.
The only exception is the nuclear recoil correction. It practically does
not depend on the nuclear charge number $Z$, since the linear $Z$ scaling in
Eq.~(\ref{eqrec}) is compensated by the increase of the nuclear mass $M$ with
$Z$. As a consequence, the recoil effect is completely negligible for high-
and medium-$Z$ ions, but turns into one of the dominant corrections for $Z<10$.

For most ions, the uncertainty of the theoretical prediction is roughly 30\% of
the Bohr-Weisskopf effect; it can be immediately estimated from
Fig.~\ref{fig:total}. For very light ions, however, there is an additional uncertainty
due to the unknown relativistic recoil effect. Its contribution
can be estimated by multiplying
the nonrelativistic recoil correction plotted on Fig.~\ref{fig:total} by
the factor of $(\Za)^2$. We observe that for very light ions, the relativistic
recoil is the dominant source of error in theoretical predictions.

\section{Conclusion}
\label{sec:5}

In this work we have performed an {\em ab initio} calculation of the
nuclear magnetic shielding in hydrogen-like ions with inclusion of relativistic, nuclear, and
QED effects. The uncertainty of our theoretical predictions for the nuclear
magnetic shielding constant defines [according to Eq.~(\ref{eq4b})] the
precision to which the nuclear magnetic dipole moments can be determined from
experiments on the $g$-factors of hydrogen-like ions. It can be concluded
from Table~\ref{tab:total} and
Fig.~\ref{fig:total} that the
present theory permits determination of nuclear magnetic moments with
fractional accuracy ranging from $10^{-9}$ in the case of
$^{17}$O$^{7+}$ to $10^{-5}$ for $^{209}$Bi$^{82+}$.

For most hydrogen-like ions, the dominant source of error in the theoretical predictions is
the Bohr-Weisskopf effect. Since
this effect cannot be accurately calculated at present, we conclude that
the theory of the nuclear magnetic shielding has reached the
point where the uncertainty due to nuclear-structure effects impedes
further progress.
For very light ions, however, the dominant theoretical error
comes from the unknown relativistic recoil effect, whose calculation might be
a subject of future work.

\section*{Acknowledgement}

Stimulating discussions with K.~Blaum and G.~Werth are gratefully
acknowledged. V.~A.~Y.~and Z.~H.~were supported by the Alliance Program of the
Helmholtz Association (HA216/EMMI). K.P.~acknowledges support by NIST
Precision Measurement Grant PMG 60NANB7D6153.

%
%
\begin{table*}[htb]
\caption{Individual contributions to the self-energy correction to the nuclear
  magnetic shielding, for the point nucleus, in terms of the function
$D_{\rm SE}$ defined by Eq.~(\ref{eqDZa}).
 \label{tab:se}}
\begin{ruledtabular}
\begin{tabular}{c......}
 $Z$ &\multicolumn{1}{c}{po} &
                     \multicolumn{1}{c}{vr,hfs} &
                                   \multicolumn{1}{c}{vr,zee} &
                                          \multicolumn{1}{c}{d.ver} &
                                               \multicolumn{1}{c}{der} &
                                                       \multicolumn{1}{c}{total}  \\
\hline\\[-7pt]
 10 &  -9.x584 &   6.x095 &   9.x180 & -49.x722 &  43.x523 &    -0.x508\,(100)  \\
 14 &  -4.x741 &   1.x711 &   4.x321 & -21.x048 &  19.x047 &    -0.x710\,(15)  \\
 16 &  -3.x571 &   0.x705 &   3.x140 & -14.x718 &  13.x655 &    -0.x789\,(9)  \\
 20 &  -2.x217 &  -0.x405 &   1.x761 &  -7.x848 &   7.x782 &    -0.x927\,(4)  \\
 26 &  -1.x277 &  -1.x115 &   0.x772 &  -3.x473 &   3.x983 &    -1.x110\,(2)  \\
 32 &  -0.x858 &  -1.x407 &   0.x292 &  -1.x644 &   2.x333 &    -1.x283\,(1)  \\
 40 &  -0.x624 &  -1.x580 &  -0.x043 &  -0.x586 &   1.x315 &    -1.x519\,(1)  \\
 45 &  -0.x575 &  -1.x643 &  -0.x171 &  -0.x267 &   0.x975 &    -1.x681\,  \\
 54 &  -0.x595 &  -1.x744 &  -0.x335 &   0.x023 &   0.x622 &    -2.x029\,  \\
 60 &  -0.x672 &  -1.x825 &  -0.x423 &   0.x112 &   0.x487 &    -2.x321\,  \\
 70 &  -0.x929 &  -2.x028 &  -0.x571 &   0.x175 &   0.x355 &    -2.x999\,  \\
 82 &  -1.x612 &  -2.x508 &  -0.x812 &   0.x191 &   0.x284 &    -4.x457\,(2)  \\
 83 &  -1.x701 &  -2.x569 &  -0.x839 &   0.x192 &   0.x281 &    -4.x636\,(1)  \\
 92 &  -3.x011 &  -3.x400 &  -1.x174 &   0.x198 &   0.x280 &    -7.x107\,(2)  \\
\end{tabular}
\end{ruledtabular}
\end{table*}

%
%
\begin{table}[htb]
\caption{Vacuum-polarization correction to the magnetic shielding, for the
  extended nucleus, in terms of the function $D_{\rm VP}$  defined by Eq.~(\ref{eq:vp}).
 \label{tab:vp}}
\begin{ruledtabular}
\begin{tabular}{cccc}
 $Z$ &\multicolumn{1}{c}{po} &
                     \multicolumn{1}{c}{mag} & \multicolumn{1}{c}{total}  \\
\hline\\[-7pt]
 10 &    0.118  &    0.110 &    0.228 \\
 14 &    0.135  &    0.121 &    0.256 \\
 16 &    0.144  &    0.126 &    0.271 \\
 20 &    0.164  &    0.137 &    0.302 \\
 26 &    0.200  &    0.155 &    0.355 \\
 32 &    0.242  &    0.175 &    0.417 \\
 40 &    0.314  &    0.206 &    0.520 \\
 45 &    0.369  &    0.228 &    0.597 \\
 54 &    0.500  &    0.275 &    0.775 \\
 60 &    0.618  &    0.315 &    0.933 \\
 70 &    0.891  &    0.398 &    1.289 \\
 82 &    1.449  &    0.546 &    1.996 \\
 83 &    1.512  &    0.562 &    2.074 \\
 92 &    2.227  &    0.727 &    2.954
\end{tabular}
\end{ruledtabular}
\end{table}

\begin{table*}
\caption{Individual contributions to the shielding constant $\sigma \times
  10^6$ for selected hydrogen-like ions.
\label{tab:total}}
\begin{ruledtabular}
  \begin{tabular}{l.....}
 &
\multicolumn{1}{c}{$^{17}$O$^{7+}$}
             &  \multicolumn{1}{c}{$^{43}$Ca$^{19+}$}
                         &\multicolumn{1}{c}{$^{73}$Ge$^{31+}$}
                                  &\multicolumn{1}{c}{$^{131}$Xe$^{53+}$}
                                         &\multicolumn{1}{c}{$^{209}$Bi$^{82+}$}  \\
    \hline\\[-5pt]
Leading        &   143.3x127      &   375.9x60      &   657.x93      &  1461.x6       &  4112x    \\
QED            &    -0.0x026\,(2) &    -0.1x03\,(15)&    -0.x59\,(8) &    -4.x1\,(0.8)&   -30x\,(7) \\
Bohr-Weisskopf &    -0.0x013\,(4) &    -0.0x61\,(18)&    -0.x54\,(16)&    -8.x2\,(2.5)&   -42x\,(13)\\
Quadrupole     &    -0.0x007\,(1) &    -0.0x18      &    -0.x42      &     6.x9\,(0.1)&     7x    \\
Recoil         &    -0.0x120      &    -0.0x15      &    -0.x02      &     0.x0       &     0x    \\
Total          &   143.2x960\,(5) &   375.7x63\,(24)&   656.x36\,(18)&  1456.x3\,(2.6)&  4046x\,(15)\\
  \end{tabular}
\end{ruledtabular}
\end{table*}

\appendix

\section{Angular integrals}
\label{app:angular}

Throughout this work, we repeatedly used the following result for the matrix element
of the hfs and Zeeman interaction
\begin{align} \label{app:eq1}
\lbr \kappa_1\mu_1|\frac{(\hr\times\balpha)_q}{r^{a}} |\kappa_2\mu_2\rbr
  &\ = (-1)^{j_2+\mu_2}\,
  \nonumber \\ & \times
C_{j_2\,-\mu_2, j_1\mu_1}^{1q}\,P^{(-a)}(n_1,n_2)\,,
\end{align}
where $\hr = \bfr/|\bfr|$, $C_{j_1m_1,j_2m_2}^{jm}$ is the Clebsch-Gordan coefficient, and
\begin{eqnarray}  \label{app:eq2}
 P^{(\alpha)}(n_1,n_2) = \frac{-\kappa_1-\kappa_2}{\sqrt{3}}\,C_1(-\kappa_2,\kappa_1)\,R^{(\alpha)}_{n_1n_2}\,.
\end{eqnarray}
The radial integral $R^{(\alpha)}$ is defined by Eq.~(\ref{eqrad}) and the
angular coefficient $C_1(\kappa_a,\kappa_b)$ is the reduced matrix element of the
normalized spherical harmonics, see, e.g., Eq.~(C10) of
Ref.~\cite{yerokhin:99:pra}.
An important particular case is $\kappa_1=\kappa_2=-1$ and
$\mu_1=\mu_2=\nicefrac12$, in which case the above formulas reduce to
\begin{eqnarray}  \label{app:eq3}
\lbr -1\,\nicefrac12|\frac{(\hr\times\balpha)_0}{r^{a}} |-1\,\nicefrac12\rbr
  = -\frac23\, R^{(-a)}_{ab}\,.
\end{eqnarray}

The basic angular integrals $K_i$ needed for the evaluation of the hfs vertex
contribution are defined by
\begin{widetext}
\begin{subequations}
\begin{align}
\frac{3i}{4\pi} \int d\hp_1\, d\hp_2\, & \ F(\xi)\,
     \chi^{\dag}_{\kappa_a \half}(\hp_1)\, [\hp_1\times \bsigma]_0\,
      \chi_{\kappa_b \half}(\hp_2)
= \int_{-1}^1 d\xi\,F(\xi)\,
        K_1(\kappa_a,\kappa_b)\,,
\\
\frac{3i}{4\pi} \int d\hp_1\, d\hp_2\, & \ F(\xi)\,
     \chi^{\dag}_{\kappa_a \half}(\hp_1)\, [\hp_2\times \bsigma]_0\,
      \chi_{\kappa_b \half}(\hp_2)
= \int_{-1}^1 d\xi\,F(\xi)\,
        K_1^{\prime}(\kappa_a,\kappa_b)\,,
\\
\frac{3i}{4\pi} \int d\hp_1\, d\hp_2\, & \ F(\xi)\,
     \chi^{\dag}_{\kappa_a \half}(\hp_1)\, [\hp_1\times \hp_2]_0\,
      \chi_{\kappa_b \half}(\hp_2)
= \int_{-1}^1 d\xi\,F(\xi)\,
        K_2(\kappa_a,\kappa_b)\,,
\end{align}
\end{subequations}
where $F(\xi)$ is an arbitrary function and $\xi = \hp_1\cdot \hp_2$.
The integrals over all angles except for $\xi$ in the above equations
are evaluated analytically, as described in Appendix of Ref.~\cite{yerokhin:10:sehfs}.
The results relevant for this work are
\begin{eqnarray}
        \begin{array}{llll}
K_1(-1,1) = -\xi, & K_1(1,-1) = 1, & K_1(-1,-2) = -\frac{\displaystyle
  \xi}{\displaystyle \sqrt{2}},
& K_1(1,2)=\frac{\displaystyle -1+3\xi^2}{\displaystyle 2\sqrt{2}},\\[2pt]
K_1^{\prime}(-1,1) = -1, & K_1^{\prime}(1,-1) = \xi, & K_1^{\prime}(-1,-2) = -\frac{\displaystyle 1}{\displaystyle \sqrt{2}}, &
    K_1^{\prime}(1,2) = \frac{\displaystyle \xi}{\displaystyle \sqrt{2}}, \\[2pt]
K_2(-1,-1) = 0, & K_2(1,1) = \frac{\displaystyle -1+\xi^2}{\displaystyle 2}, &
K_2(-1,2) = 0,
& K_2(1,-2) = \frac{\displaystyle -1+\xi^2}{\displaystyle 2\sqrt{2}}.
        \end{array}
\end{eqnarray}

The basic angular integrals $A_i$ needed for the evaluation of the Zeeman vertex
contribution are given by
\begin{align}
A_1(\kappa_a,\kappa_b) &\ = \int d\hp\, \chi^{\dagger}_{\kappa_a \half}(\hp)\,
 \sigma_0\, \chi_{\kappa_b \half}(\hp)\,, \\
A_2(\kappa_a,\kappa_b) &\ = \int d\hp\, \chi^{\dagger}_{\kappa_a \half}(\hp)\,
 i\,[\bsigma\times \hp]_0\, \chi_{\kappa_b \half}(\hp)\,, \\
A_3(\kappa_a,\kappa_b) &\ = \int d\hp\, \chi^{\dagger}_{\kappa_a \half}(\hp)\,
 i\,[\bsigma\times \vec{\nabla}_{\Omega}]_0\, \chi_{\kappa_b \half}(\hp)\,, \\
A_4(\kappa_a,\kappa_b) &\ = \int d\hp\, \chi^{\dagger}_{\kappa_a \half}(\hp)\,
 i\,[\hp\times \vec{\nabla}_{\Omega}]_0\, \chi_{\kappa_b \half}(\hp)\,.
\end{align}
These integrals are evaluated by using the standard Racah algebra.
The results relevant for this work are
\begin{eqnarray}
        \begin{array}{llll}
A_1(-1,-1) = 1, & A_1(1,1) = -1/3, & A_1(-1,2) = 0, & A_1(1,-2) = -2\sqrt{2}/3, \\[2pt]
A_2(-1,1) = 2/3,& A_2(1,-1)=-2/3,  & A_2(-1,-2) = \sqrt{2}/3, & A_2(1,2)=-\sqrt{2}/3, \\[2pt]
A_3(-1,1) = 4/3,& A_3(1,-1) = 0,  & A_3(-1,-2) = 2\sqrt{2}/3, & A_3(1,2)=-\sqrt{2},\\[2pt]
A_4(-1,-1) = 0, & A_4(1,1) = -2/3, & A_4(-1,2) = 0, & A_4(1,-2) = -\sqrt{2}/3.
        \end{array}
\end{eqnarray}
\end{widetext}


\end{document}